\documentclass[12pt]{article}
\usepackage{epsf,epsfig,float,amssymb,latexsym}
\usepackage{graphics,psfrag}
\usepackage{amsmath}
\usepackage{rotating} 

\newlength{\MySep}
\newcommand {\Emptie} {\rule{0ex}{1ex}}

\newcommand{\beq}{\begin{equation}}
\newcommand{\eeq}[1]{\label{#1}\end{equation}}
\newcommand{\beqa}{\begin{eqnarray}}
\newcommand{\eeqa}[1]{\label{#1}\end{eqnarray}}
\newcommand{\eeqan}{\end{eqnarray}}
\newcommand{\CR}{\nonumber \\ }

\newcommand {\CosXXX}      {\frac{\sqrt{3}}{2}}

\newcommand {\CosXXXthird} {\frac{\sqrt{3}}{6}}
\newcommand {\CosXLV}      {\frac{\sqrt{2}}{2}} 
\newcommand {\CosXLVhalf}  {\frac{\sqrt{2}}{4}} 
\newcommand {\TanXXX}      {\frac{\sqrt{3}}{3}} 
\newcommand {\ROSixthHalf} {\frac{\sqrt{6}}{12}} 
\newcommand {\BeNul}       {b_{0}}
\newcommand {\BeDe}        {b_{\rm D}}
\newcommand {\BeFe}        {b_{\rm F}}
\newcommand {\DeNul}       {d_{0}}
\newcommand {\DeDe}        {d_{\rm D}}
\newcommand {\DeFe}        {d_{\rm F}}
\newcommand {\DeI}         {d_{1}}
\newcommand {\DeII}        {d_{2}}
\newcommand {\DxD}         {D^{2}}
\newcommand {\DxF}         {DF}
\newcommand {\FxF}         {F^{2}} 
\newcommand {\MyHBox}      {\rule{4em}{0em}}

\newcommand {\MySepRule}   {\rule{0em}{1.425em}}
\newcommand {\MinSepRule}  {\rule[-.65em]{0em}{1em}}  
\newcommand {\RootOfSixth} {\frac{\sqrt{6}}{6}} 

\makeatletter
\def\gsim{\compoundrel>\over\sim}
\def\lsim{\compoundrel<\over\sim}
\def\compoundrel#1\over#2{\mathpalette\compoundreL{{#1}\over{#2}}}
\def\compoundreL#1#2{\compoundREL#1#2}
\def\compoundREL#1#2\over#3{\mathrel
      {\vcenter{\hbox{$\m@th\buildrel{#1#2}\over{#1#3}$}}}}
\makeatother

\begin{document}

\bigskip\bigskip

\begin{center}{{\Large\bf  
Separable potential model for\\
$K^{-}N$ interactions at low energies 
}}\end{center}

\begin{center}
{\large A.~Ciepl\'y$^{a}$, J.~Smejkal$^b$}
\end{center}

\bigskip

{\noindent
$^a${\it Nuclear Physics Institute, 250 68 \v{R}e\v{z}, Czech Republic \\}
$^b${\it Institute of Experimental and Applied Physics, 
Czech Technical University, Horsk\'{a} 3a/22,\\ 128~00~Praha~2, 
Czech Republic}
}

\begin{abstract}
\noindent 
The effective separable meson-baryon potentials are constructed to match 
the equivalent chiral amplitudes up to the second order in external meson 
momenta. We fit the model parameters (low energy constants) to the threshold 
and low energy $K^{-}p$ data. In the process, the $K^{-}$-proton bound state 
problem is solved exactly in the momentum space and the 1s level characteristics 
of the kaonic hydrogen are computed simultaneously with the available 
low energy $K^{-}p$ cross sections. The model is also used to 
describe the $\pi \Sigma$ mass spectrum and the energy dependence of the 
$K^{-}n$ amplitude.
\end{abstract}

 
\section{Introduction}
\label{sec:int}

The meson-baryon interactions at low energies have become a testing ground 
for theoretical models based on chiral symmetry. Since the pioneering works 
of Weinberg \cite{79Wei}, Gasser, Leutwyler \cite{84GLe} and others 
(see \cite{03Sch} for a comprehensive overview) 
the chiral perturbation theory has been established 
as the effective field theory of strong interactions that implements the QCD 
symmetries in a region where perturbative QCD is inapplicable. In the SU(2) sector 
the ChPT proved to be quite successful thanks to very small current masses 
of the $u$ and $d$ quarks. The smallness of the pion mass also complies well 
with its presumed origin as that of the Goldstone boson. 

The situation becomes more intriguing once we enter the strange sector. 
Especially, the treatment of the kaon-nucleon interaction at low energies 
requires a special care. Unlike the pion-nucleon interaction the $\bar{K}N$ 
dynamics is strongly influenced by the existence of the $\Lambda(1405)$ 
resonance, just below the $K^{-}p$ threshold. This means that the standard 
chiral perturbation series do not converge. Fortunately, one can use 
non-perturbative coupled channel techniques to deal with the problem and 
generate the $\Lambda(1405)$ resonance dynamically. Though such approach 
violates the crossing symmetry it has proven 
quite useful and several authors have already applied it to various low 
energy meson-baryon processes \cite{95KSW}-\cite{06Oll}. 

While the properties of the $\Lambda(1405)$ resonance have been well known 
for a long time, the nature of the resonance still remains a mystery. For many 
years it has been considered as a meson-baryon quasi-bound state coupled
to the $\pi\Sigma$ and $\bar{K}N$ channels \cite{77JDH}. 
It can also be viewed as a standard 
$qqq$ baryon \cite{78IKa} and some authors have advocated the notion that 
the resonance is a pentaquark state \cite{07Ino}. Recently, it was also 
realized that the chiral models generate two poles in the complex energy plane 
that can be assigned to the $\Lambda(1405)$ \cite{01OMe}, \cite{03JOO}. 
Although the "two poles model" may look viable and supported by the analysis 
\cite{05MOR} of the $K^{-}p \rightarrow \pi^{0}\pi^{0}\Sigma^{0}$ 
measurement \cite{04Pra} it is still not quite clear 
if (and how) this particular set of experimental data 
is compatible with the results of other experiments related to the lineshape 
of the $\Lambda(1405)$ resonance. We will come back to this point in Section 
\ref{sec:Lambda1405} while discussing the relevant pole structure 
and the $\pi \Sigma$ mass spectrum generated by our model.      

There is a plenty of experimental data on various processes initiated 
by the $K^{-}p$ interaction at low energies. The relatively 
old data on cross sections and threshold branching ratios were supplemented 
by recent measurements of strong interaction effects on the 1s level 
of kaonic hydrogen (KEK \cite{98KEK} and 
DEAR collaboration in Frascati \cite{05DEAR}). 
Although the measurements have confirmed the repulsive 
character of the $K^{-}p$ strong interaction at threshold the DEAR 
values of the strong interaction shift and width of the 1s level look 
at odds with the $K^{-}p$ scattering length extrapolated from the scattering 
measurements. In this report we present our analysis of the situation 
and discuss the compatibility of the kaonic hydrogen measurement with 
other data. The novelty of our approach lies in exact calculation of 
the $K^{-}p$ bound state properties instead of relying on the approximate 
Deser-Trueman relation \cite{DTr} (or its 
version modified to include the isospin effects and electromagnetic corrections 
\cite{04MRR}). Our separable potential model can also serve as a viable 
alternative to the N/D scheme based on a dispersion relation for the inverse 
of the T-matrix that employs techniques and language common in high energy 
physics. The separable potentials are well suited for any few body calculations 
that involve the $\bar{K}N$ interactions at low energies. Particularly, 
it should not be difficult to adapt our model for the Faddeev type calculations 
of the $\bar{K}NN - \pi \Sigma N$ system \cite{07SGM}.  

A brief account of our work was already given in 
\cite{07CSm}. Here we expand the letter not only by providing more 
details on our approach but include also a discussion of the $\Lambda(1405)$ 
resonance spectrum and present our results for the $K^{-}n$ scattering 
amplitude. Since the scope 
of the article is wider than in our previous work and we apply our model 
to a broader interval of the $\bar{K}N$ energies (specifically below the 
$\bar{K}N$ threshold) we put additional constraints on the model parameters 
and present completely new fits to the data with those constraints in effect. 

Our main aim remains a simultaneous description of both the 1s level 
kaonic bound state and the available experimental data for the $K^{-}p$ 
initiated processes. The characteristics of kaonic hydrogen 1s level are 
computed precisely with the same effective chiral potentials that are 
used to calculate the properties of $K^{-}p$ induced reactions.  As we showed
in \cite{07CSm}, the direct computation of the kaonic hydrogen characteristics 
is becoming necessary in view of the experimental precision expected 
in the SIDDHARTA measurement executed in Frascati \cite{07Cur}. At the same 
time our results for the $K^{-}n$ elastic amplitude are relevant to 
the measurement of kaonic deuterium performed by the same collaboration. 

The article is organized as follows. First we outline the method we use to 
calculate the quantities observed in low energy $K^{-}p$ interactions, then 
we present the chiral Lagrangian used to derive our effective meson-baryon 
potentials. Main part of the paper is given in Section \ref{sec:results} 
where we present 
and discuss the results obtained in our fits to the $K^{-}p$ experimental 
data and proceed with the analysis of the $\pi \Sigma$ mass spectrum and 
the $K^{-}n$ interactions. Our conclusions are briefly summarized 
in the last section. 

\section{Method outlined}
\label{sec:for}

We developed a precise method of computing the meson-nuclear bound
states in momentum space. The method was already applied to pionic 
atoms \cite{94CMa} and its multichannel version was used to calculate the 1s level
characteristics of pionic hydrogen \cite{96CMa}. Here we just 
remark that our approach is based on the construction of the Jost matrix 
and involves the solution of the Lippman-Schwinger equation for the transition 
amplitudes between various channels. Bound states in a specific 
channel then correspond to zeros of the determinant of the Jost matrix 
at (or close to) the positive part of the imaginary axis in the complex 
momentum plane. The zeros are searched for iteratively by means of the Mueller 
algorithm. If only the point-like 
Coulomb potential is considered in the $K^{-}p$ channel the method reproduces 
the well known Bohr energy of the 1s level with a precision better 
than $0.1$~eV. The inclusion of the leading electromagnetic corrections, 
the charge finite sizes and vacuum polarization effects, gives 
an attractive energy shift $\Delta E_{FS+VP}(1s) = -20.2$ eV.

The necessary ingredient needed to calculate the impact of strong interaction 
on $K$-atomic energy levels is the kaon-nuclear optical potential. In the case 
of kaonic hydrogen and multiple channels it means the potential matrix. 
We follow the approach of Ref.~\cite{95KSW} and construct the strong
interaction part of the potential matrix as effective transition amplitudes 
that give the same (up to the order $\mathcal{O}(q^2)$ of the external meson  
momenta) s-wave scattering lengths as are those derived from the underlying 
chiral Lagrangian. While the authors of Ref.~\cite{95KSW} restricted themselves 
only to the first six meson-baryon channels that are open 
at the $\bar{K}N$ threshold we employ all ten coupled meson-baryon channels: 
$\pi^{0}\Lambda$, $\pi^{0}\Sigma^{0}$, $\pi^{-}\Sigma^{+}$, $\pi^{+}\Sigma^{-}$, 
$K^-p$, $\bar{K}^{0}n$, $\eta \Lambda$, $\eta \Sigma^0$, $K^0\Xi^0$, and $K^+ \Xi^-$. 
We order the channels according to their threshold energies and will refer 
to them (and index them) in this particular order. 
As we already mentioned in the Introduction the $\Lambda(1405)$ resonance 
does not enter as a separate field but it is generated dynamically 
by solving coupled Lippman-Schwinger equations with the input potential matrix. 

The strong interaction potential matrix is given in the separable form 
\beq
V_{ij}(k,k')=\sqrt{\frac{1}{2E_i}\frac{M_i}{\omega_i}}\:g_{i}(k)
\:\frac{C_{ij}}{f^2} \:g_{j}(k')\:\sqrt{\frac{1}{2E_j}\frac{M_j}{\omega_j}}
, \quad g_{j}(k)=\frac{1}{1+(k/ \alpha_{j})^2}
\eeq{eq:poten}
in which the momenta $k$ and $k'$ refer to the meson-baryon c.m. system 
in the $i$ and $j$ channels, respectively. The kinematical factors 
$\sqrt{M_j/(2 E_j \omega_j)}$ guarantee a proper relativistic flux 
normalization with the meson energy $E_j$ and the baryon mass and energy 
$M_j$ and $\omega_j$, all taken in the c.m. system of channel $j$.
The off shell form factors $g_{j}(k)$ introduce the inverse range 
radii $\alpha_{j}$ that characterize the radius of interactions in 
various channels. Finally, the parameter $f \simeq 100$ MeV 
(a value between the empirical pion and kaon decay constants) stands for 
the pseudoscalar meson decay constant in the chiral limit 
and the coupling matrix $C_{ij}$ 
is determined by chiral SU(3) symmetry and includes terms up to the second 
order in the meson c.m.~kinetic energies. The details on the underlying 
chiral Lagrangian and on the couplings $C_{ij}$ will be given in the following 
section.

The potential of Eq.~(\ref{eq:poten}) is used not only when solving the bound 
state problem but we also implement it in the standard Lippman-Schwinger 
equation and compute the low energy $\bar{K}N$ cross sections and branching 
ratios from the resulting transition amplitudes. Our LS equation 
for the s-wave coupled channel T-matrix and the separable potential 
(\ref{eq:poten}) can be written in a purely algebraic form,
\beq
t_{ij} = v_{ij} + \displaystyle\sum_{n} v_{in}\: I_{n}\: t_{nj} \;\;\; ,
\eeq{eq:LS}
where we introduced the notation $T_{ij} = t_{ij}g_{i}g_{j}$ and 
$V_{ij} = v_{ij}g_{i}g_{j}$ (the obvious dependence on kinematical variables 
is not shown here for simplicity). In vacuum, the integral $I_n$ can be  
evaluated analytically, 
\beq
I_{n} = 2\mu_{n} \int \frac{d^{3}l}{(2\pi)^{3}}\: 
\frac{g^{2}_{n}(l)}{k^{2}_{n}-l^{2}+i\epsilon} = 
- \frac{\mu_{n}}{2\pi}\: \frac{(\alpha_{n}+ik_{n})^{2}}{2\alpha_{n}}\: 
g^{2}_{n}(k_{n}) \;\;\; .
\eeq{eq:In} 
Here $k_{n}$ stands for the on-shell meson-baryon relative momenta in the 
intermediate channel $n$ and $\mu_{n}$ denotes the "reduced mass" 
of the system, $\mu_{n} = E_{n}\omega_{n}/(E_{n} +\omega_{n})$.
The nonrelativistic scattering amplitude $f_{ij}$ for the transition from 
channel $j$ to channel $i$ is then simply obtained by solving the system 
of algebraic equations (\ref{eq:LS}) and by using the relation 
$f_{ij} = -\sqrt{\mu_{i}\mu_{j}}\, T_{ij}/(2\pi)$. The observable quantity,
the total s-wave cross section for the transition, is given by the standard 
formula, 
\beq
\sigma_{ij} = 4\pi\: \frac{k_{i}}{k_{j}} \: |f_{ij}|^{2}
\;\;\; .
\eeq{eq:xsect} 

The reader should note that our approach differs from the recently more popular 
on-shell N/D scheme based on the Bethe-Salpeter equation, unitarity relation 
for the inverse of the $T$-matrix and on the dimensional regularization 
of the scalar loop integral \cite{01OMe}. Though the difference is only 
a technical one it has consequences. 
The advantage of our method is that the off-shell form 
factors are parameterized by means of the inverse range radii which have 
a better physical meaning than the subtraction constants appearing 
due to the regularization procedure used in the "inverse $T$-matrix approach". 
In principle the off-shell effects can be incorporated in the latter model too. 
However, in our approach they appear quite naturally with no additional effort. 
On the other hand the use of Bethe-Salpeter equation and quantum field techniques 
makes the other model more attractive it terms of completely relativistic dynamics 
while we restrict ourselves only to relativistic treatment of the kinematical 
variables. This restriction is fully justified in the region of low 
and intermediate energies that are the subject of our work.

\section{Chiral Lagrangian}

In this section we briefly outline the effective chiral Lagrangian 
that is based on the $SU(3)_{\rm L} \otimes SU(3)_{\rm R}$ chiral symmetry 
and reflects the symmetries of QCD. 
It describes the coupling of the pseudoscalar meson octet 
($\pi$, $K$, $\bar{K}$, $\eta$) 
to the ground state baryon octet ($N$, $\Lambda$, $\Sigma$, $\Xi$). 
Following Ref.~\cite{95KSW} we consider the first two orders 
(in terms of the external meson momenta and quark masses) 
of the Lagrangian density, 
\begin{equation}
  {\cal L} = {\cal L}^{(1)} + {\cal L}^{(2)}.
\end{equation}
The leading order reads
\beq
  {\cal L}^{(1)} = Tr( \overline{\Psi}_B(i \gamma_\mu D^\mu - M_0) \Psi_B ) +
           F \, Tr( \overline{\Psi}_B \gamma_\mu \gamma_5 [ u^\mu ,\Psi_B]) +
           D \, Tr( \overline{\Psi}_B \gamma_\mu \gamma_5 \{ u^\mu ,\Psi_B \})
\eeq{eq:lagr1}
where the covariant derivative is given by
\begin{equation}
   D^\mu \Psi_B = \partial^\mu \Psi_B + 
   [{1 \over {8f^2}} [\phi,\partial^\mu \phi],\Psi_B]+\dots
\end{equation}
and the axial matrix operator is
\begin{equation}
u^{\mu} = -\frac{1}{2f}\partial^\mu \phi + \dots \; .
\end{equation}
The meson and (heavy) baryon fields are represented by the matrices $\phi$ 
and $\Psi_B$, respectively. Further, $M_0$ is the baryon mass 
in the chiral limit and the constants $F$ and $D$ are the $SU(3)$ 
axial vector couplings. 

The leading order (linear in the external meson four-momentum $q$) 
of Eq.~(\ref{eq:lagr1}) is the current algebra, or the Weinberg-Tomozawa, term.
In addition, the Lagrangian ${\cal L}^{(1)}$ gives rise to s-wave meson-baryon 
amplitudes at order $q^2$ (and higher). They appear due to relativistic 
corrections to the covariant derivative term and due to the Born graphs 
terms that originate from the axial coupling part of ${\cal L}^{(1)}$.  
These $q^2$ pieces add to the relevant s-wave terms of the second order 
Lagrangian, 
\begin{eqnarray}
{\cal L}^{(2)} &=& b_D Tr(\overline{B} \lbrace \chi_+ , B \rbrace)+b_F
 Tr(\overline{B} \lbrack \chi_+ , B \rbrack)  + b_0 Tr(\overline{B} B)
 Tr(\chi_+) \CR
        & + & d_D Tr(\overline{B} \lbrace (u^2 + (v \cdot u)^2),B \rbrace) +
        d_F Tr(\overline{B} \lbrack (u^2 + (v \cdot u)^2),B \rbrack) \CR
    & + & d_0 Tr(\overline{B} B) Tr(u^2 + (v \cdot u)^2)   
    \label{eq:lagr2} \\
    & + & d_1 (Tr(\overline{B} u_\mu) Tr(u^\mu B) + Tr(\overline{B} 
    (v \cdot u)) Tr((v \cdot u) B) ) \CR
    & + & d_2 Tr(\overline{B} (u_\mu B u^\mu + (v \cdot u) B (v \cdot u)))
                  + \cdots \nonumber
\end{eqnarray}
which, at the tree level, also generates contributions of the order $q^2$.
We have denoted
\begin{equation}
  \chi_+ = - {1 \over {4 f^2 }} \lbrace \phi, \lbrace \phi, \chi\rbrace \rbrace
\end{equation}
and $v^{\mu}$ is the baryon four-velocity, $v^{\mu} = (1,0,0,0)$ for baryon 
at rest. The $\chi$ matrix introduces explicit chiral symmetry breaking and 
is proportional to the quark mass matrix with only diagonal elements 
not equal to zero. In the isospin symmetry limit, the diagonal
elements are $(m^2_\pi , m^2_\pi , 2m^2_K - m^2_\pi )$. 

\begin{figure}
\caption{The Feynman diagrams relevant for the s-wave meson-baryon interaction.}
\centering
\begin{tabular}{cccc}
\includegraphics[width=0.2\textwidth]{feynps.1} & 
\includegraphics[width=0.2\textwidth]{feynps.2} &  
\includegraphics[width=0.2\textwidth]{feynps.3} &
\includegraphics[width=0.2\textwidth]{feynps.4} \\
$\mathcal{O}(q^1)$ contact & $\mathcal{O}(q^2)$ contact & 
direct s-term & crossed u-term
\end{tabular}
\label{fig:graphs}
\end{figure}

When the effective meson-baryon potentials (\ref{eq:poten}) are constructed 
to match the Born amplitudes generated by the chiral Lagrangian the low energy 
constants (the couplings at various terms in the Lagrangian) combine 
to the couplings $C_{ij}$ that bind the considered meson-baryon states. 
Thus, the chiral symmetry of meson-baryon interactions is reflected 
in the structure of the $C_{ij}$ coefficients derived directly 
from the Lagrangian. The general structure of the couplings reads as
\beqa
  C_{ij}
    & =
    & - C_{i j}^{\rm (WT)} \,
        \frac{\:\! E'_{i} + E'_{j} \!\:}{4}
      \:+\: C_{i j}^{(mm)} \:\!
        \left( m_{i}^{2} + m_{j}^{2} \right)
      \:+\: C_{i j}^{(\chi {\rm b})} \:\!
        \left( m_{K}^{2} - m_{\pi}^{2} \right)
       +\: C_{i j}^{(EE)} \,
        E_{i}^{\Emptie} E_{j}^{\Emptie}+    \CR[.5em]
  \Emptie
    &
    & \Emptie 
      \:+\: C_{i j}^{(s)} \,
        \frac{\:\! E_{i}^{\Emptie} E_{j}^{\Emptie}}%
          {2 M_{0}^{\Emptie} \!\:}
      \: \Emptie +\: C_{i j}^{(u)} \,
        \frac{1}{\:\! 3 M_{0}^{\Emptie} \!\:} \,
        \left(
          2 m_{i}^{2} + 2 m_{j}^{2}
            + \frac{\:\! m_{i}^{2} m_{j}^{2} \!\:}%
              {E_{i}^{\Emptie} E_{j}^{\Emptie}}
            - \textstyle{\frac{7}{2}}
              E_{i}^{\Emptie} E_{j}^{\Emptie}
        \right) \; ,
\eeqa{eq:Cij}
where the primed meson energies $E'_{j}$ include the relativistic 
correction, $E'_{j} = E_{j} + (E_{j}^{2}-m^{2}_{j})/(2M_{0})$, with $m_{j}$ 
denoting the meson mass in the channel $j$.
The terms marked by the superscripts "WT", "s" and "u" correspond to 
the leading Weinberg-Tomozawa contact interaction and to the direct 
and crossed Born amplitudes, respectively. The remaining parts contribute 
to the contact interaction in the next-to-leading ({\it i.e.} $q^{2}$) order. 
For brevity we show the pertinent graphs in Figure \ref{fig:graphs} 
which we take from Ref.~\cite{06BMN}. The terms $(mm)$ and $(\chi {\rm b})$ 
both appear due to explicit breaking of the chiral symmetry. In fact, what 
we denoted as the $(\chi {\rm b})$ term represents even the violation  
of the vector $SU(3)_{V}$ symmetry that is reduced to the isospin $SU(2)_{V}$ 
symmetry, {\it i.e.} the flavor symmetry of the $SU(3)$ octets is broken 
by this term. 

The actual composition of the coefficients $C_{i j}^{\rm (.)}$ is given 
in the Appendices. We note that the coefficients for the first six channels 
coupled to the $K^{-}p$ system that are open at the $\bar{K}N$ threshold 
were already published in \cite{95KSW} 
while in Appendix A we show the complete tables for all ten considered 
channels. The couplings $C_{i j}$ can be related to their counterparts 
used in the alternative approach based on the chiral Lagrangian that 
is manifestly invariant to Lorentz transformations. The relation was derived 
in Ref.~\cite{00FMM} for the case of the $SU(2)$ chiral symmetry. The derivation 
for the $SU(3)$ case is more complex and goes beyond the scope of the present 
work. In principle, the approaches based on both formulations of the chiral 
Lagrangian should give the same results for physical observables. However, 
this is true only when one sums up all orders of the infinite series 
of the relevant Feynman diagrams (all orders in $q$), not once we restrict 
ourselves to a given perturbative order (here $q^{2}$). This means that our 
results provided in the next section may (to a reasonable extent) differ 
from those achieved with the alternative formulation of the Lagrangian.

\section{Results}
\label{sec:results}

In this section we closely follow the line presented in our letter \cite{07CSm} 
and show the results of our fits to the available low energy $\bar{K}N$ 
experimental data. While in \cite{07CSm} we aimed our analysis only at 
the kaonic hydrogen characteristics, the $K^{-}p$ threshold branching ratios 
and at the cross sections of $K^{-}p$ initiated reactions, here we also 
include the position of the $\Lambda(1405)$ resonance observed in the 
$\pi \Sigma$ mass spectrum. Additionally, we also present an analysis 
of the $K^{-}n$ scattering amplitude and discuss the effects due to 
breaking of isospin symmetry. 

\subsection{$\bar{K}N$ data fits}
\label{sec:fits}

The three precisely measured threshold branching ratios \cite{81Mar} are
\beqa
\gamma & = & \frac{\sigma(K^-p\rightarrow \pi^+\Sigma^-)}
             {\sigma(K^-p\rightarrow \pi^-\Sigma^+)}=2.36\pm 0.04~, \CR
R_c    & = & \frac{\sigma(K^-p\rightarrow \hbox{charged particles})}
             {\sigma(K^-p\rightarrow \hbox{all})}=0.664\pm0.011~, 
             \label{eq:rates} \\
R_n    & = & \frac{\sigma(K^-p\rightarrow \pi^0\Lambda)}
             {\sigma(K^-p\rightarrow \hbox{all neutral states})}
             = 0.189\pm 0.015~. \nonumber
\end{eqnarray}
They impose quite tight constraints on any model applied to the $\bar{K}N$ 
interactions at low energies. 

The cross sections of $K^{-}p$ initiated 
reactions are not determined so accurately, thus they do not restrict the fits 
so much. We consider only the experimental data taken at the kaon laboratory 
momenta $p_{LAB} = 110$ MeV (for the $K^- p$, $\bar{K^0}n$, $\pi^{+} \Sigma^{-}$,
$\pi^{-} \Sigma^{+}$ final states) and at $p_{LAB} = 200$ MeV (for 
the same four channels plus $\pi^0 \Lambda$ and $\pi^{0} \Sigma^{0}$). 
Although some authors include in their fits the experimental cross sections 
at all available kaon momenta we feel that such approach unduely magnifies 
the importance of this particular set of data at expense of all other 
measurements that are not represented by so many data points.     
Anyway, our results show that the inclusion of the cross section data taken 
at many kaon momenta is not necessary since the fit at just $1-2$ points 
fixes the cross section magnitude and the energy dependence is reproduced 
nicely by the model. 

We apply the same philosophy to the measured $\pi \Sigma$ mass distribution 
and fit only the position of the peak at 1395 MeV instead of fitting 
the complete measured spectra. Again, this appears to be quite sufficient 
as we will see in Section \ref{sec:Lambda1405}. In a manner of Ref.~\cite{01OMe} 
we assume that the $\pi \Sigma$ mass distribution originates from a generic 
$s$-wave isoscalar source which couples to the $\bar{K} N$ and $\pi \Sigma$ $I=0$ 
states. Since the measured event distribution is not normalized, only the ratio 
of the relevant couplings $r = r_{\bar{K} N} / r_{\pi \Sigma}$ is of significance. 
In other words, we assume that the observed $\pi \Sigma$ spectrum complies 
with the prescription
\beq
dN_{\pi\Sigma}/dE \sim 
\bigg| T_{\pi\Sigma, \pi\Sigma}(I=0) + r_{KN/ \pi \Sigma}\: T_{\pi\Sigma, \bar{K}N}(I=0) \bigg|^2 
p_{\pi\Sigma}
\eeq{eq:pisig}
where the ratio $r_{KN/ \pi \Sigma}$ is energy independent. In general, the ratio 
should be a complex number but (to further simplify the matter) we consider 
only real values in our fits. Though the reality may not be so simple 
the ansatz looks appropriate for simulating the dynamics of the $\Lambda(1405)$ 
resonance. 

Finally, we include the DEAR results \cite{05DEAR} on the strong interaction 
shift $\Delta E_N$ and the width $\Gamma$ of the 1s level in kaonic hydrogen:
\beq
\Delta E_{N}(1s)= (193 \pm 43)  \hbox{ eV,}\;\;\;
\Gamma(1s) = (249 \pm 150) \hbox{ eV~.}
\eeq{Katom}  
Thus, we end up with a total of 16 data points in our fits.

The parameters of our model are: a) the couplings of the chiral 
Lagrangian which enter the coefficients $C_{ij}$, b) the inverse 
range radii $\alpha_{j}$ that provide the off-shell behavior of the potentials 
$V_{ij}$ of Eq.~(\ref{eq:poten}), c) the ratio $r_{KN/ \pi \Sigma}$ 
determining the relative coupling of the $\bar{K} N$ 
and $\pi \Sigma$ channels to the $\Lambda(1405)$ resonance. 
Apparently, the number of parameters is too large, so it is desirable 
to fix some of them prior to performing the fits. First, the axial couplings 
$D$ and $F$ were already established in the analysis of semileptonic hyperon 
decays \cite{99Rat}, $D = 0.80$, $F = 0.46$ ($g_{A} = F + D = 1.26$). 
Then, we set the couplings $b_D$ and $b_F$ to satisfy the approximate 
Gell-Mann formulas for the baryon mass splittings, 
\beqa
M_{\Xi} - M_N            &=& - 8 b_F (m_K^2 -m_\pi^2) \CR
M_{\Sigma} - M_{\Lambda} &=& \frac{16}{3} b_D (m_K^2 -m_\pi^2)~, 
\eeqan
which gives $b_D = 0.064$ GeV$^{-1}$ and $b_F = -0.209$ GeV$^{-1}$. 
Similarly, we determine the coupling $b_0$ and the baryon chiral mass 
$M_0$ from the relations for the pion-nucleon sigma term $\sigma_{\pi N}$ 
and the proton mass,
\beqa
\sigma_{\pi N} &=& -2 m_\pi^2(2b_0+b_D+b_F)  ~, \CR
M_p            &=& M_0-4m_K^2( b_0+ b_D-b_F)-2 m_\pi^2 (b_0+2b_F) ~.
\eeqa{eq:sigma}
Since the value of the pion-nucleon $\sigma$-term is not well established we 
enforce four different options, $\sigma_{\pi N} =$ (20--50) MeV, which 
cover the interval of the values considered by various authors.
Finally, we reduce the number of the inverse ranges $\alpha_{j}$ 
to only five: $\alpha_{KN}$, $\alpha_{\pi \Lambda}$, 
$\alpha_{\pi \Sigma}$, $\alpha_{\eta \Lambda /\Sigma}$, 
$\alpha_{K \Xi}$. This leaves us with 12 free parameters: the five inverse 
ranges, the meson-baryon chiral coupling $f$, the ratio $r_{KN/ \pi \Sigma}$ 
of Eq.~(\ref{eq:pisig}) and five more low energy constants 
from the second order chiral Lagrangian 
denoted by $d_D$, $d_F$, $d_0$, $d_1$, and $d_2$. 

The number of the second order couplings can be reduced even further 
since the pertinent Lagrangian terms are not completely independent. 
Thanks to the Cayley-Hamilton identity any of the Lorentz invariants contributing 
to the second order Lagrangian (\ref{eq:lagr2}) with the $d$-couplings  
can be expressed as a linear combination of the other four invariants. 
This feature is reflected in the SU(3) 
chiral coefficients $C_{ij}$ which are invariant under the transformation
\beq
\begin{array}{lcccl}
d_D & \longrightarrow & d_D' & = & d_D + \Delta     \\
d_F & \longrightarrow & d_F' & = & d_F              \\
d_0 & \longrightarrow & d_0' & = & d_0 - \Delta /2  \\
d_1 & \longrightarrow & d_1' & = & d_1 - \Delta     \\ 
d_2 & \longrightarrow & d_2' & = & d_2 + \Delta 
\end{array}
\eeq{eq:d-shift}
for any real $\Delta$. In other words, one of the couplings (besides $d_F$) 
can be set to zero. While in our previous work we did not use this property 
here we set $d_2 = 0$ and fit only the remaining low energy constants. 

Our results are summarized in Tables \ref{fits} - \ref{tab:alphas}. The first table 
shows the results of our $\chi^{2}$ fits compared with the relevant experimental 
data. The resulting $\chi^{2}$ per data point indicate satisfactory 
fits. It is worth noting that their quality and the computed values 
do not depend much on the exact value of the $\sigma_{\pi N}$ term. 
Tables \ref{param} and \ref{tab:alphas} show the fitted parameters 
of the chiral Lagrangian and the inverse range parameters $\alpha_j$. 
The last rows in the tables compare our values with those determined 
in Ref.~\cite{95KSW} (with the $d$-couplings shifted according 
to Eq.~(\ref{eq:d-shift}) to satisfy the condition $d_2 = 0$). 
The reader may note that the fitted parameters differ significantly 
from those given in our earlier report \cite{07CSm}. The main reason 
for this is that in the present fits we decided to restrict the values 
of the inverse ranges $\alpha_{j}$ while in Ref.~\cite{07CSm} we allowed for 
practically unrestricted region of the pertinent parameter space. 
Since the parameters $\alpha_{j}$ represent inverse ranges of meson-baryon 
interactions it seems natural to restrict their values from below 
by the mass of the lightest meson, the pion. Additionally, we want 
to avoid any unphysical resonances that might appear due to possible
poles in the off-shell form factors $g_{j}(k)$ for $k^{2}+\alpha_{j}^{2}$.
For the on-shell momenta the poles appear at the cms energies below 
the respective meson-baryon thresholds. The requirement that such poles 
may not lie at the real energy axis leads to the condition 
$\alpha_{j}>m_{j}$. In the energy region relevant for the present work,
$1300$ MeV $\lsim \sqrt{s} \lsim 1500$ MeV, one can disregard 
the unphysical poles that are sufficiently far from the considered 
energy interval and use slightly weakened restrictions on $\alpha_{j}$.
We have found it sufficient to restrict the search of the inverse range 
parameters to the following intervals: $\alpha_{\pi \Lambda} > 150$ MeV, 
$\alpha_{\pi \Sigma} > 150$ MeV, $\alpha_{\bar{K} N} > 350$ MeV, 
$\alpha_{\eta \Lambda /\Sigma} > 500$ MeV 
and $150$ MeV $< \alpha_{K \Xi} < 300$ MeV or $\alpha_{K \Xi} > 500$ MeV.         
Of course, we also checked that if we lift the restrictions placed on 
the $\alpha$ parameters we are able to reproduce our earlier results. 
In fact, the use of the Cayley-Hamilton identity 
[incorporation of Eq.~(\ref{eq:d-shift})] does not alter the results 
reported in Ref.~\cite{07CSm} at all and the introduction of a new fitted 
quantity, the peak position of the $\pi \Sigma$ mass distribution 
results in only minor changes of the original parameter sets.
We also noted that tuning the peak position of the $\pi \Sigma$ 
mass spectrum has no effect on the other $\bar{K}N$ data.
One can simply fit all available $\bar{K}N$ data first (as we did in 
Refs.~\cite{07CSm}) and then shift the position of the peak by adjusting 
the parameter $r_{KN/ \pi \Sigma}$. We will come back to this point 
in the following section and show that one can get quite reasonable 
description of the $\pi \Sigma$ spectrum even for our previous parameter sets. 

In general, we conclude that the fits are not affected much by 
the inclusion of the $\pi \Sigma$ mass spectrum and by the Cayley-Hamilton identity 
enforced on the $d$-couplings. The physically motivated restrictions applied 
to the inverse ranges $\alpha_{j}$ lead to different local $\chi^{2}$ minima 
but the quality of the fits remains good. Although 
our new fits have slightly higher values of $\chi^{2}/N$ than those reported 
in Ref.~\cite{07CSm} the new parameter constraints guarantee that the 
computed amplitudes do not suffer from any unphysical resonances in the 
interval of energies from $1300$ to $1500$ MeV.  

Finally, we remind the reader that the parameter $b_0$ and the 
baryon mass in the chiral limit were not fitted to the data and are given 
in the second and third column of Table \ref{param} only to visualize their 
respective values corresponding to the selected $\sigma_{\pi N}$ term. 
The $\pi N$ isospin-even scattering 
length $a_{\pi N}^{+}$ shown in the fourth column of Table \ref{param} was 
not included in our fits either but we feel that its presentation 
is important and deserves some comments. 

\begin{table}
\caption{The fitted $\bar{K}N$ threshold data}
\begin{center}
\begin{tabular}{ccccccc}
$\sigma_{\pi N}$ [MeV] & $\chi^{2}/N$ & $\Delta E_{N}$ [eV] & $\Gamma$ [eV]& $\gamma$ & $R_c$ & $R_n$ \\ \hline
 20           & 1.33  & 214     & 718      & 2.368   & 0.653 & 0.189 \\
 30           & 1.29  & 260     & 692      & 2.366   & 0.655 & 0.188 \\
 40           & 1.35  & 195     & 763      & 2.370   & 0.654 & 0.191 \\
 50           & 1.37  & 289     & 664      & 2.366   & 0.658 & 0.192 \\ \hline
 exp          &   -   & 193(43) & 249(150) & 2.36(4) & 0.664(11) & 0.189(15)
\end{tabular}
\end{center}
\label{fits}
\end{table}

\begin{table}
\caption{Chiral Lagrangian parameters ($b_0$ and $d$'s in $1/$GeV):}
\begin{center}
\begin{tabular}{ccccccccc}
$\sigma_{\pi N}$ [MeV] &  $M_0$ [MeV] & $a_{\pi N}^{+}$ [$m_{\pi}^{-1}$]& $f$ [MeV] & $d_0$ & $d_D$ & $d_F$ & $d_1$ \\ \hline
 20                    &  997 & -0.009 & 111.0 & -0.108 & -0.446 & -0.834 & 0.540  \\
 30                    &  864 & -0.001 & 109.1 & -0.450 &  0.026 & -0.601 & 0.235  \\
 40                    &  729 & -0.007 & 114.5 & -0.492 & -0.635 & -0.788 & 0.616  \\
 50                    &  594 &  0.002 & 107.6 & -1.043 &  0.229 & -0.478 & 0.161  \\ \hline
 27 (Ref.~\cite{95KSW}) & 910 & -0.002 &  94.5 & -0.71  &  0.38  & -0.43  & -0.34 
\end{tabular}
\end{center}
\label{param}
\end{table}

The low energy constants involved in our fits should also be  
constrained by other observables calculated within the framework 
of ChPT involving the same meson-baryon Lagrangian. The spectrum of baryon 
masses and the $\pi N$ isospin-even scattering length may come to one's mind 
in this respect. The later quantity to order $q^3$ is given by \cite{93BKM}:
\beq
  a^+_{\pi N} = {{1} \over {4 \pi (1+m_\pi /M_N)}}\biggl[ {{m^2_\pi} \over
            {f^2}} \biggl(-2b_D - 2b_F - 4b_0 + d_D + d_F + 2 d_0 -
                   {{g^2_A} \over {4 M_N}} \biggr) 
              +  {{3 g^2_A m^3_\pi} \over {64 \pi f^4}}  \biggr]~.
\eeq{a_piN}
Since the experimental value of $a^+_{\pi N}$ is practically consistent with 
zero, $a_{0+}^+=-(0.25\pm 0.49) \cdot 10^{-2}\,m_\pi^{-1}$ \cite{99Sch}, 
the $d$-parameters should combine to give a negative contribution 
that cancels the positive one due to the $b$ terms and the $q^{3}$
correction represented by the last term in Eq.~(\ref{a_piN}). 
As a smaller $\sigma_{\pi N}$ term means a smaller absolute 
value of the negative parameter $b_0$ (and hence a smaller positive 
contribution due to the $b_0$ term in $a_{\pi N}^{+}$)  
the computed $\pi N$ scattering length should become negative 
for too low $\sigma_{\pi N}$ terms. Considering the fact that many other 
authors ({\it e.g.} \cite{95KSW} or \cite{06Oll}) include the $a_{\pi N}^{+}$ 
value directly in their fits, it is interesting that our fits 
aimed purely at the $\bar{K}N$ interactions allow for so good reproduction 
of the $\pi N$ quantity. One can also view the agreement of our model 
with the vanishing value of the $a_{\pi N}^{+}$ as an independent 
confirmation that the model complies with the chiral symmetry.

Although we have performed fits for $\sigma_{\pi N}=20$ MeV 
there is no reason to believe that the $\sigma_{\pi N}$ value should 
be so small. In fact, such a small value leads to a negative strangeness 
content in the proton, 
\begin{equation}
    {{<p|\overline{s}s|p>} \over  {<p|\overline{u}u + \overline{d}d |p>}} =
                      {{b_0+b_D-b_F} \over {2b_0+b_D+b_F}},
\end{equation}
when one considers only the contributions to the order of $q^{2}$. 
We also feel that the value $\sigma_{\pi N}=50$ MeV represents 
rather a maximal limit for any considerations and a feasible choice 
should be around $\sigma_{\pi N} =$ (30--40) MeV.

We have also tried to perform fits with the $b$ parameters taken from the
analysis of the baryon mass spectrum \cite{97BMe} and with only the current 
algebra (Weinberg-Tomozawa) term contributing to the $C_{ij}$ coefficients 
(the approach adopted in Ref.~\cite{98ORa}). Unfortunately, we were not 
able to achieve satisfactory results in those cases. Our best fits performed  
without the second order terms gave $\chi^{2}/N \gsim 3$. Thus, it looks that 
the low energy constants derived in the analysis of baryon masses are not 
suitable in the sector of meson-baryon interactions and that the inclusion of 
the $q^{2}$ terms is necessary for a good description of the $\bar{K}N$ data. 
A comprehensive discussion of the importance of various second 
order $q^{2}$ contributions to the computed observables was given 
in Ref.~\cite{05BNW}.

\begin{table}
\caption{Inverse range parameters (in MeV):}
\begin{center}
\begin{tabular}{cccccc}
$\sigma_{\pi N}$ [MeV] & $\alpha_{\pi \Lambda}$ & $\alpha_{\pi \Sigma}$ & 
  $\alpha_{KN}$ & $\alpha_{\eta \Lambda /\Sigma}$ & 
  $\alpha_{K \Xi}$ \\ \hline
  20     & 226 & 579 & 625 & 917 & 260 \\ 
  30     & 291 & 601 & 639 & 568 & 151 \\ 
  40     & 219 & 640 & 638 & 936 & 226 \\ 
  50     & 345 & 600 & 608 & 507 & 152 \\ \hline
 27~\cite{95KSW} & 300 & 450 & 760 &  -  &  -  
\end{tabular}
\end{center}
\label{tab:alphas}
\end{table}

The inverse range parameters shown in Table \ref{tab:alphas} are in line with 
our expectations. The values corresponding to the open channels $\bar{K}N$, 
$\pi \Lambda$ and $\pi \Sigma$ seem to be well determined and show only 
a moderate dependence on the adopted value of the $\sigma_{\pi N}$ term. 
In general, the ranges obtained for the open channels correspond to the 
t-channel exchanges that are believed to dominate the interactions. 
The restrictions we applied on the inverse ranges in the present work 
do not affect the fitted values of $\alpha_{j}$ in the open channels.
On the other hand the range of interactions in the closed channels is not
well defined in the fits and the fitted values $\alpha_{\eta \Lambda /\Sigma}$  
and $\alpha_{K \Xi}$ exhibit relatively large statistical errors. This 
feature also justifies our use of only one range parameter for both 
$\eta$ channels. If the inverse ranges were not constrained by any limits 
(as it was so in our previous work \cite{07CSm}) the fitted values of  
$\alpha_{\eta \Lambda /\Sigma}$ and $\alpha_{K \Xi}$ would be quite different 
from those given it the Table \ref{tab:alphas}. This indicates that 
the minima found of our $\chi^{2}$ fits differ from those found in 
the "unrestricted" fits. While working 
on the fits we also noted that the $\bar{K} N$ data prefer very small 
values of the $K\Xi$ inverse range. If there are no restrictions put on 
the $\alpha_{K\Xi}$ its value tends to get as small as $10-30$ MeV which is 
unphysical. Of course, one could advocate the slightly better fits of the 
$K^{-}p$ data without the restrictions applied to the inverse ranges 
$\alpha_{j}$ but we prefer a broader applicability of our model (to a larger 
interval of cms energies) and more meaningful values of its parameters. 
We will come back to this point in section \ref{sec:Kn} and show how this 
influences the energy dependence of the $K^{-}n$ amplitude.   

\begin{figure}
\includegraphics[width=\textwidth]{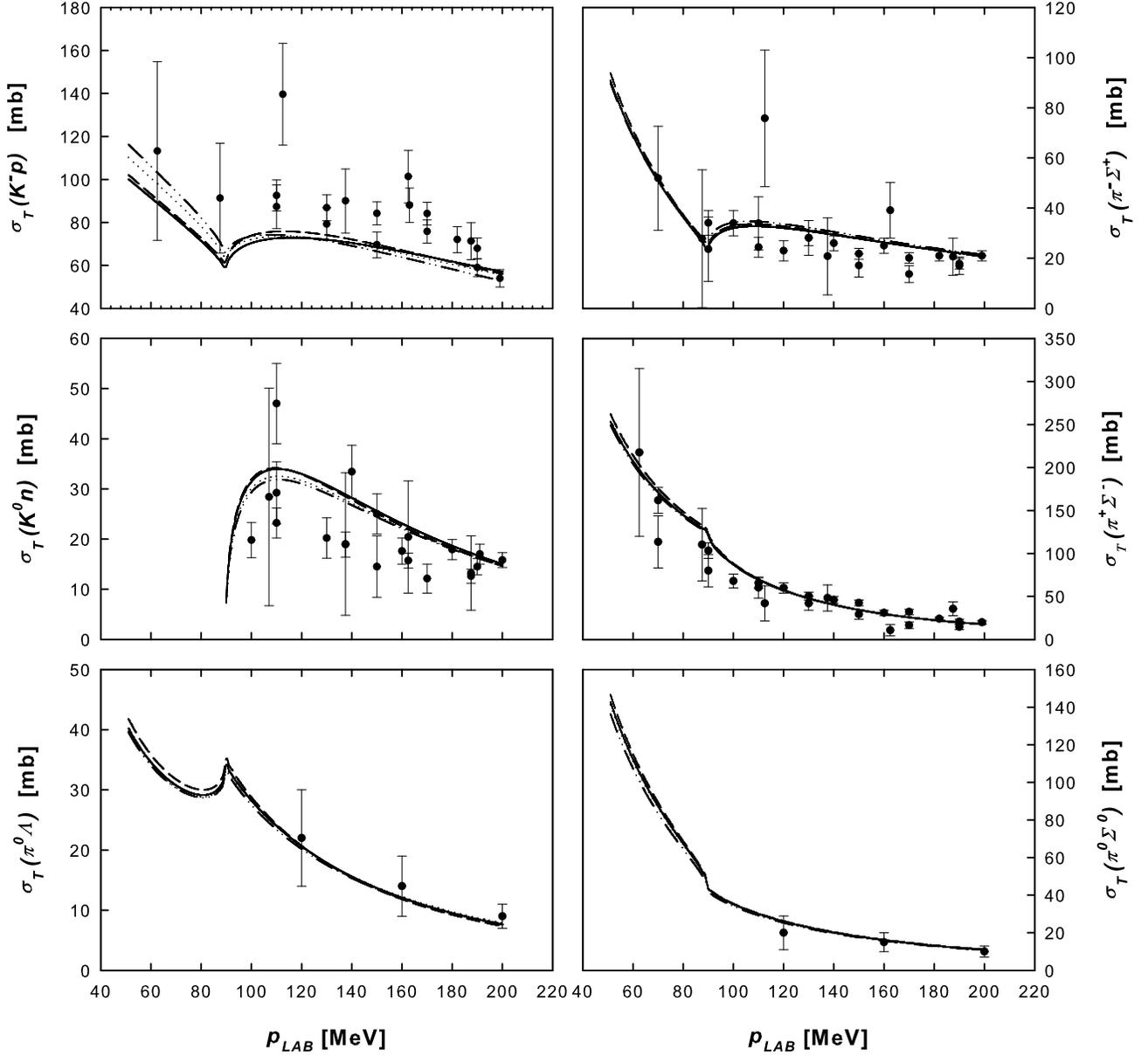} 
\caption{Total cross sections for $K^{-}p$ scattering and reactions to the 
meson-baryon channels open at low kaon laboratory momenta $p_{LAB}$. 
The experimental data are the same as those compiled in Fig.~1 
of Ref.~\cite{95KSW}. Our results obtained for $\sigma_{\pi N}=$ 20, 30, 40 
and 50 MeV are visualized by the full, dotted, dashed and dot-dashed lines, 
respectively.}
\label{fig:xsections}
\end{figure}

In Figure \ref{fig:xsections} we present the low energy $K^{-}p$ initiated cross 
sections. The results obtained for various adopted values of $\sigma_{\pi N}$ 
are practically undistinguishable with the only exception at low kaon 
momenta in the elastic channel. This observation is rather puzzling since 
the experimental cross sections are not so much restrictive as the threshold 
branching ratios. Apparently, the parameter space is flexible enough 
to accommodate the fitted values. Though we declined 
from using all experimental data in our fits and took only the data points 
available for the selected kaon laboratory momenta $p_{LAB} = 110$ MeV 
and $p_{LAB} = 200$ MeV, the description of the data is quite good. 
Specifically, we do not observe the lowering of the calculated cross sections 
in the elastic $K^{-}p$ channel reported by Borasoy et al.~\cite{05BNW} 
for their fits including the DEAR kaonic hydrogen characteristics. Though our 
$K^{-}p$ cross sections are also slightly below the experimental data 
the difference is not significant. In addition, the inclusion of 
electromagnetic corrections discussed in Ref.~\cite{05BNW} should partly 
improve the description for the lowest kaon momenta.

Finally, let us turn our attention to the calculated characteristics 
of the 1s level in kaonic hydrogen. The strong interaction 
energy shift of the 1s level in kaonic hydrogen is reproduced well but we 
were not able to get a satisfactory fit of the 1s level energy width as our 
results are significantly larger than the experimental value. This result is 
in line with the conclusions reached by Borasoy, Meissner and Nissler 
\cite{06BMN} on the basis of their comprehensive analysis of the $K^{-}p$ 
scattering length from scattering experiments. However, when considering 
the interval of three standard deviations and also the older KEK 
results \cite{98KEK} (which give less precise but larger width) we cannot 
conclude that kaonic hydrogen measurements contradict the other low energy 
$\bar{K}N$ data. We hope the new SIDDHARTA experiment performed in Frascati 
will clarify the situation concerning the kaonic hydrogen characteristics. 
In view of its expected precision it becomes necessary to solve 
the $K^{-}p$ bound state problem exactly (as we do here) rather 
than relate the $K$-atomic characteristics to the $K^{-}p$ scattering length. 
We have shown \cite{07CSm} that the difference may be as large 
as about $10\%$, which is more than the anticipated precision 
of the SIDDHARTA measurement.

\subsection{$\Lambda(1405)$ resonance}
\label{sec:Lambda1405}

As we mentioned in the Introduction the origin and structure of the 
$\Lambda(1405)$ resonance observed in the $\pi \Sigma$ mass spectrum 
are an actively pursued topic. The coupled channel meson-baryon 
models based on chiral symmetry generate the resonance dynamically and
it appears that there are two poles in the complex energy plane that may 
contribute to the observed spectrum \cite{01OMe}. This recent discovery 
has stimulated both the theoretical debates as well as experimental 
efforts aiming at a better understanding of the $\Lambda(1405)$ structure. 
 
\begin{figure}[h]    
\begin{center}
\scalebox{0.85}{\includegraphics{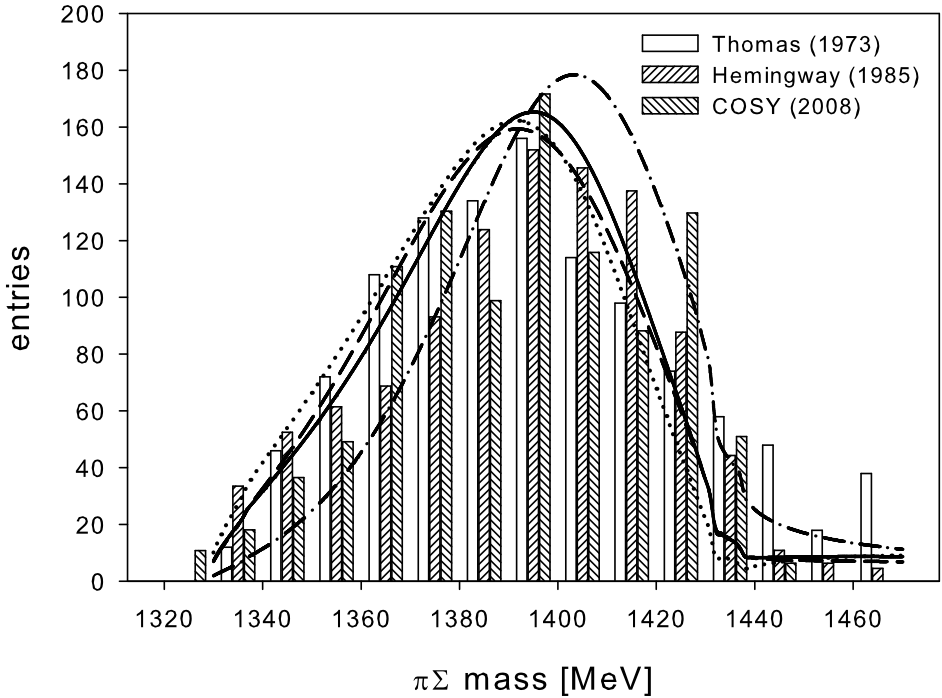}}
\end{center}
\caption{The $\pi \Sigma$ mass distribution. Our results are compared with 
the experimental data taken from Refs.~\cite{03Tho}, \cite{84Hem} and 
\cite{08Zyc}. The full line was obtained for the 
parameter set of the present work, the dashed line for the one of 
Ref.~\cite{07CSm}, both related to $\sigma_{\pi N} = 40$ MeV. See the text  
for explanation on the dotted and dash-dotted lines.}
\label{fig:pisig}
\end{figure}

The Fig.~\ref{fig:pisig} visualizes the $\pi \Sigma$ mass distribution 
computed for the parameter sets related to $\sigma_{\pi N} = 40$ MeV.  
In addition to the distribution obtained for the present fit (and represented 
by a full line in the figure) we also show (dashed line in the figure) the 
$\pi \Sigma$ spectrum generated for the pertinent parameter set 
of Ref.~\cite{07CSm} and $r_{KN/ \pi \Sigma}=1$. It peaks at $1291$ MeV 
and we would need $r_{KN/ \pi \Sigma}=2.6$ to shift the spectrum to peak 
at $1395$ MeV. The shape of the spectrum is not much affected by tuning 
the parameter $r_{KN/ \pi \Sigma}$ within reasonable limits. Just for 
a reference we also show the spectra obtained by assuming that the 
$I=0$ resonance originates only from the $\pi \Sigma$ channels 
($r_{KN/ \pi \Sigma} = 0$, dotted line in Fig.~\ref{fig:pisig}) 
or that it is formed exclusively from the $\bar{K}N$ channels 
($1/r_{KN/ \pi \Sigma} = 0$, dot-dashed line). These two lines represent 
a kind of boundaries on the shape and peak position of the spectra 
in a situation when the low energy constants are fixed at the values 
obtained in our current fit for $\sigma_{\pi N} = 40$ MeV. 
Similar picture can also be drawn for the other 
choices of $\sigma_{\pi N}$.  

The experimental data shown in Fig.~\ref{fig:pisig} come from three different 
measurements \cite{03Tho}, \cite{84Hem}, \cite{08Zyc}, 
all exhibiting a prominent structure around 1400 MeV. 
As the observed spectra are not normalized we have rescaled the original 
data as well as our computed distributions to give 1000 events in the chosen 
energy interval (from 1330 to 1440 MeV). The three measurements give  
$\pi \Sigma$ distributions that look mutually compatible.
We have not included in the figure  
the $K^{-}p \rightarrow \Sigma^{0} \pi^{0} \pi^{0}$ data measured 
by the Crystal Ball Collaboration \cite{04Pra} as they yield a slightly different 
distribution with a peak structure around 1420 MeV. The two identical pions 
in the final state of the later reaction complicate a comparison with 
the other experiments, so we find it questionable to relate our computed 
lineshape to the one observed in \cite{04Pra} without employing fully 
the dynamics of the particular reaction as it was done in \cite{05MOR}. 

The values of $r_{KN/ \pi \Sigma}$ obtained in our fits (and presented in the 
Table \ref{tab:poles}) are compatible with similar findings  by other authors 
\cite{03JOO}, \cite{05BNW}. Since the magnitude of $r_{KN/ \pi \Sigma}$ 
is of the order of one it looks that both the $\bar{K}N$ and the $\pi \Sigma$ 
states contribute to the $I=0$ resonance identified with $\Lambda (1405)$ 
with a comparable strength. In other words, the inclusion of the initial 
$\bar{K}N$ channels in the model driven by Eq.~(\ref{eq:pisig}) 
is important. This is fully in line with the well 
known fact that the $\Lambda(1405)$ resonance does couple strongly to 
the $K^{-}p$ state. Unfortunately, the experimental data 
are not precise enough to distinguish between various values of 
$r_{KN/ \pi \Sigma}$ which is demonstrated in Fig.~\ref{fig:pisig} 
by comparing our best fit results with those generated for the boundary 
values of $r_{KN/ \pi \Sigma}$. Since a good description of the spectrum can 
already be achieved without its inclusion in the fits one may argue that 
the dynamics of the chiral model is fixed by the threshold (and low energy) 
observables of $K^{-}p$ interactions. However, the data clearly prefer 
a positive sign of $r_{KN/ \pi \Sigma}$, {\it i.e.}~a constructive interference 
of the contributions provided by the $\pi \Sigma$ and $\bar{K}N$ channels 
to the resonance. For the negative values of $r_{KN/ \pi \Sigma}$ 
the peak moves to energies lower than the one obtained at the 
$r_{KN/ \pi \Sigma} = 0$ boundary and the computed spectrum no longer 
matches the experimental one.     

\begin{table}
\caption{The complex energies of the poles relevant to the $I=0$ resonance.}
\begin{center}
\begin{tabular}{cccccc}
$\sigma_{\pi N}$ [MeV] & $r_{KN/ \pi \Sigma}$ 
                      & Re $z_1$ [MeV] & Im $z_1$ [MeV] & Re $z_2$ [MeV] & Im $z_2$ [MeV] 
                              \\ \hline
         20    & 1.28 &  1395          & -49            &   1456         &  -77  \\
         30    & 1.32 &  1398          & -51            &   1441         &  -76  \\
         40    & 0.37 &  1401          & -41            &   1519         & -112  \\
         50    & 0.54 &  1406          & -39            &   1436         & -138     
\end{tabular}
\end{center}
\label{tab:poles}
\end{table}
   
The interest in the $\pi \Sigma$ mass distribution has arisen since 
discovering that the chiral meson-baryon dynamics generates two poles 
in the complex energy plane that can be related to the $\Lambda(1405)$ 
resonance. In the Table \ref{tab:poles} we show the positions of the poles 
generated by our model. They appear on the unphysical Riemann sheet 
accessed when crossing the real axis between the thresholds 
of the $\pi \Sigma$ and the $\bar{K}N$ channels. The position of the lower 
(with the lower value of the real part of the complex energy) pole 
is moreless stable and does not depend much on the choice of the parameter set. 
Its complex energy $z \approx 1400 - {\rm i}\:45$ MeV can clearly be associated 
with the observed $\pi \Sigma$ mass spectrum. On the other hand, the higher 
(in terms of Re $z$) pole is located further from the real axis 
and its position vary with the chosen parameter set. We have also noted that 
parameter sets obtained for various local $\chi^{2}$ minima lead to different 
positions of this pole even if they correspond to the same choice 
of the $\pi N$ sigma term.

Interestingly, neither of the two poles is located so close to the real 
energy axis as other authors claim. This feature can be explained 
by a different parametrization of our model. It was already shown 
by Borasoy et al.~\cite{05BNW} that the second pole moves away from 
the real axis when the second order terms are included in the chiral 
Lagrangian. Our observations confirm this. When we performed a fit 
(for $\sigma_{\pi N} = 40$ MeV) with the next-to-leading order terms neglected 
we located the poles at $z_{1} = (1368 -{\rm i}\: 42)$ MeV  and 
$z_{2} = (1441 -{\rm i}\: 22)$ MeV. Although the second pole remains 
above the $\bar{K}N$ threshold while other authors observe it about 10 MeV 
below the threshold, the closeness of the pole to the real axis seems to be 
related to the omission of the next-to-leading order corrections in the chiral 
Lagrangian. We also noted that our fits with interaction restricted only 
to the Weinberg-Tomozawa term require very large values of the parameter 
$r_{KN/ \pi \Sigma}$, typically $r_{KN/ \pi \Sigma} \gsim 10$. This means 
that in such a case the resonance observed in the $\pi \Sigma$ mass spectrum 
couples much stronger to the $\bar{K}N$ channels than to the $\pi \Sigma$ 
ones.

Though the quality of the fit is much worse without the second order terms 
(we got $\chi^2 /N = 3.1$ in the case mentioned here), most of the $K^{-}p$ 
data are still reproduced quite well. On the other hand 
Hyodo and Weise \cite{08HWe}, who locate the $\bar{K}N$ quasibound state 
at $\sqrt{s} \simeq 1420$ MeV, use a parametrization that is not 
suitable for a description of all relevant $\bar{K}N$ data, specifically 
they do not reproduce (as one can check in \cite{03HNJ}) the precise threshold 
rates, Eq.~(\ref{eq:rates}). Borasoy et al.~\cite{05BNW} do fit all relevant 
$K^{-}p$ data including the threshold rates, however their pole at 1420 MeV 
moves away from the real axis (and to lower energies) when they include 
the DEAR data in their fits. When they compromise the DEAR data with those 
from $K^{-}p$ reactions they get the pole quite close to where we see it.  
Therefore, it looks plausible that the remaining differences in exact 
localization of the poles can be attributed either to model specifics 
or to the fact that the models do not reproduce all observed experimental 
data on the same footing. In a comment made by one of us and 
A.~Gal \cite{08CGa} we also showed that the position of the poles 
can change drastically when playing with 
the meson-baryon channel couplings. Thus, it should not be surprising 
that different parametrizations of the chiral model lead to different 
pole positions. 
   
The shape of the $\pi \Sigma$ mass spectrum is determined by the positions 
of the two $I=0$ poles and by the relative couplings of the $\pi \Sigma$ 
and $\bar{K}N$ states to the poles (the parameter $r_{KN/ \pi \Sigma}$ 
in our model). It is obvious that the observed spectrum does not resemble 
a typical Breit-Wigner resonance. While this can be attributed to an 
interplay of two resonances that are relatively close to the real axis 
\cite{03JOO} we can explain the $\pi \Sigma$ distribution without any resonance 
that sits in a vicinity of the real axis. In our model the observed structure 
is not of the Breit-Wigner type simply because there is no pole sufficiently 
close to the real axis. We hope that new results from experiments dedicated to 
exploration of the $\Lambda(1405)$ structure will be able to distinguish 
between those two pictures.

\subsection{$K^{-}n$ amplitude}
\label{sec:Kn}   

Once the low-energy constants of the chiral Lagrangian, Eqs.~(\ref{eq:lagr1}) 
and (\ref{eq:lagr2}), and the inverse ranges of meson-baryon 
interactions are fixed to the $K^{-}p$ data the model provides us with 
predictions for other interactions of the meson octet with the baryonic one. 
Here we discuss our results for the meson-baryon systems with total charge $Q=-1$, 
specifically for the $K^{-}n$ amplitude. In the $Q=-1$ sector the coupled 
channels are represented by the following ones:  $K^{-}\Lambda$, 
$\pi^{-}\Sigma^{0}$, $\pi^{0}\Sigma^{-}$, $K^{-}n$, $\eta \Sigma^{-}$, 
$K^{0} \Xi^{-}$ (listed and numbered according to the respective thresholds).
Exactly as in the $Q=0$ sector related to the $K^{-}p$ system we construct 
the effective potentials (\ref{eq:poten}) with the coupling matrix 
$C_{ij}$ given in Appendix B. 

\begin{figure}
\scalebox{0.92}{\includegraphics{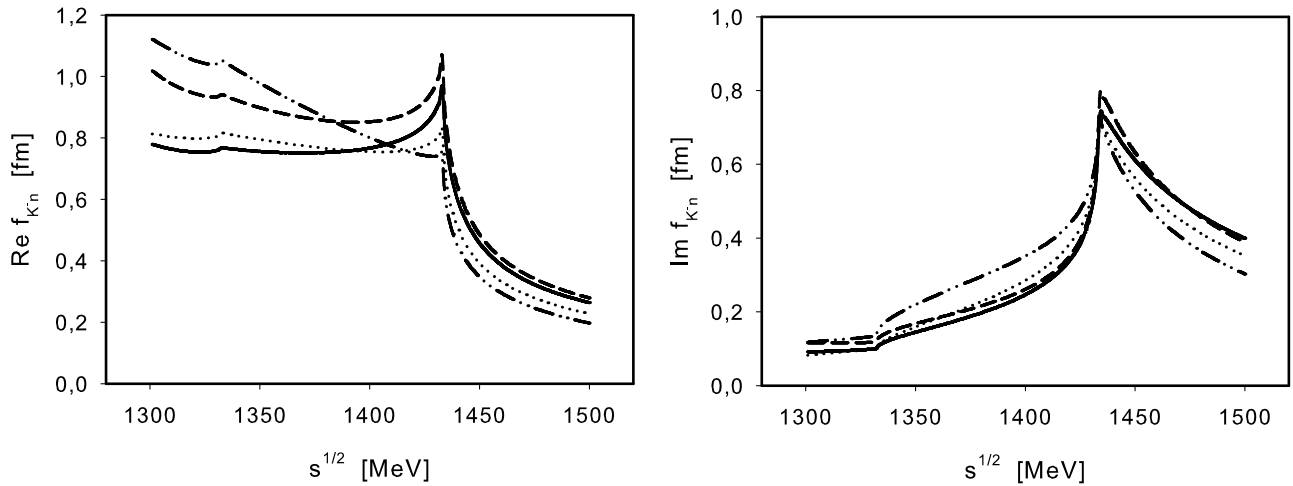}} 
\caption{The real (left panel) and imaginary (right panel) parts of the 
$K^{-}n$ amplitude. The full, dotted, dashed and dot-dashed lines 
visualize our results obtained with the parameter sets corresponding to 
$\sigma_{\pi N}$ terms set to 20, 30, 40 and 50 MeV, respectively.}
\label{fig:Kn}
\end{figure}

\begin{figure}
\scalebox{0.92}{\includegraphics{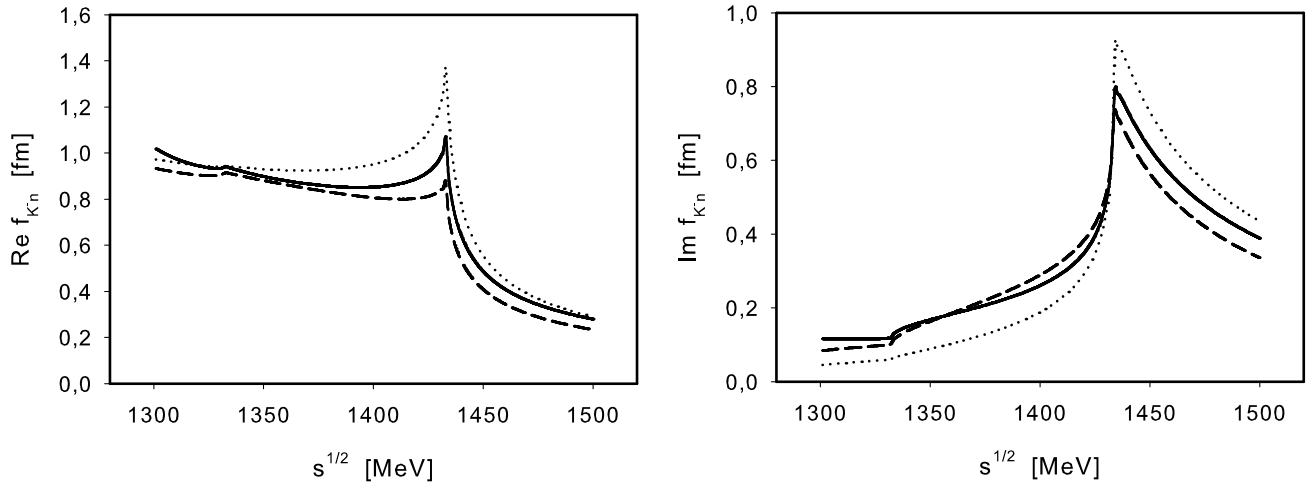}} 
\caption{The real (left panel) and imaginary (right panel) parts of the 
$K^{-}n$ amplitude. Results are presented only for the parameter set 
corresponding to $\sigma_{\pi N} = 40$ MeV. See text for more details.}
\label{fig:Kn-fits}
\end{figure}

In Figure \ref{fig:Kn} we present our results for the elastic $K^{-}n$ 
amplitude as a function of the c.m.s. energy. It is a bit surprising 
to see how much the calculated amplitudes depend on the choice of the parameter 
set (related to the value of the $\sigma_{\pi N}$ term). We have demonstrated 
that all four choices give an equivalent description of the available 
$K^{-}p$ data, so one would expect a similar feature in the $K^{-}n$ sector 
as well. This is not true either at the $\bar{K}N$ threshold or below it. 
At the $K^{-}n$ threshold the variations in the real part of the elastic 
amplitude make as much as some 30\%. It is also worth noting that for energies 
below the threshold the real part of the $K^{-}n$ amplitude 
follows a different trend than the one reported in Refs.~\cite{05BNW} and 
\cite{08HWe}. Specifically, our Re $f_{K^{-}n}$ is either a slightly 
decreasing or a moreless constant function of the energy while in \cite{05BNW} 
and \cite{08HWe} it turned out as monotonically increasing function of energy 
below the threshold. While the authors of the first paper \cite{05BNW} work 
with the physical meson and baryon masses and their approach is similar to ours 
(parameters fitted 
to $K^{-}p$ data used to compute the $K^{-}n$ amplitude) the results of 
the more recent work by Hyodo and Weise \cite{08HWe} were obtained in a model 
that adopts fully the isospin symmetry and does not aim at a realistic 
description of the $K^{-}p$ threshold branching ratios. Since the other 
authors do not incorporate off-shell effects and use a 
different formulation of the $\bar{K}N$ dynamics it is difficult to trace  
the origin of the observed differences. We have checked that the inclusion 
of the "u" terms (corresponding to Fig.~\ref{fig:graphs}d) in the calculation 
(which the other authors refrained from) leads only to a minor modification 
of the $K^{-}n$ amplitude below the threshold. We demonstrate 
the effect in Figure \ref{fig:Kn-fits} where the full line represents the 
present calculation completed with all terms included and the dotted line 
corresponds to the calculation without the "u" terms. For the real part of the 
amplitude both lines moreless coincide 
at energies above the $K^{-}n$ threshold. The dashed line represents 
the results obtained for the parameter set of Ref.~\cite{07CSm}, our previous 
fit to the $K^{-}p$ data performed without any restrictions on the inverse 
ranges $\alpha_{j}$. Since the full and dashed lines correspond to two 
different $\chi^{2}$ minima (but to the same choice of the $\sigma_{\pi N}$) 
the lineshape variations give an idea of the theoretical uncertainies 
inherent in our description of the $K^{-}n$ amplitude.

\begin{figure}
\begin{center}
\scalebox{0.8}{\includegraphics{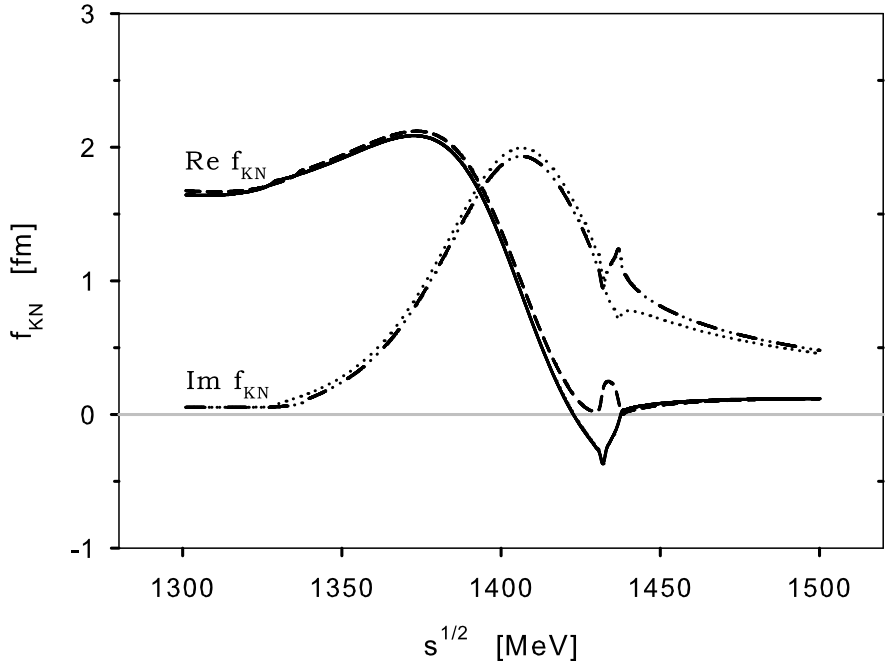}}
\end{center} 
\caption{The energy dependence of the $K^{-}p$ and $\bar{K}^{0}n$ elastic 
amplitudes. The real and imaginary parts of the amplitudes are shown in 
the same figure with the full and dotted lines used for the $K^{-}p$ 
amplitude and the dashed and dot-dashed lines used for the $\bar{K}^{0}n$ 
amplitude, respectively.}  
\label{fig:KpxKn}
\end{figure}

In principle, the elastic $K^{-}n$ amplitude can be related to 
the isovector parts of the $\bar{K}N$ amplitudes 
obtained in the $Q=0$ sector, {\it i.e.}~for the coupled channel model used to 
describe the $K^{-}p$ data. However, the physical meson and baryon masses 
break the isospin symmetry and the threshold energies of different $\bar{K}N$ 
channels are different too. Thus, one should be careful when using 
the isospin relations at or near the $\bar{K}N$ thresholds. To demonstrate 
the ambiguity, in Figure \ref{fig:KpxKn} we present a comparison of the 
elastic $K^{-}p$ and $\bar{K}^{0}n$ amplitudes. Although both amplitudes 
have the same isospin content, their behaviour at threshold energies is 
quite different.

\begin{figure}
\scalebox{0.92}{\includegraphics{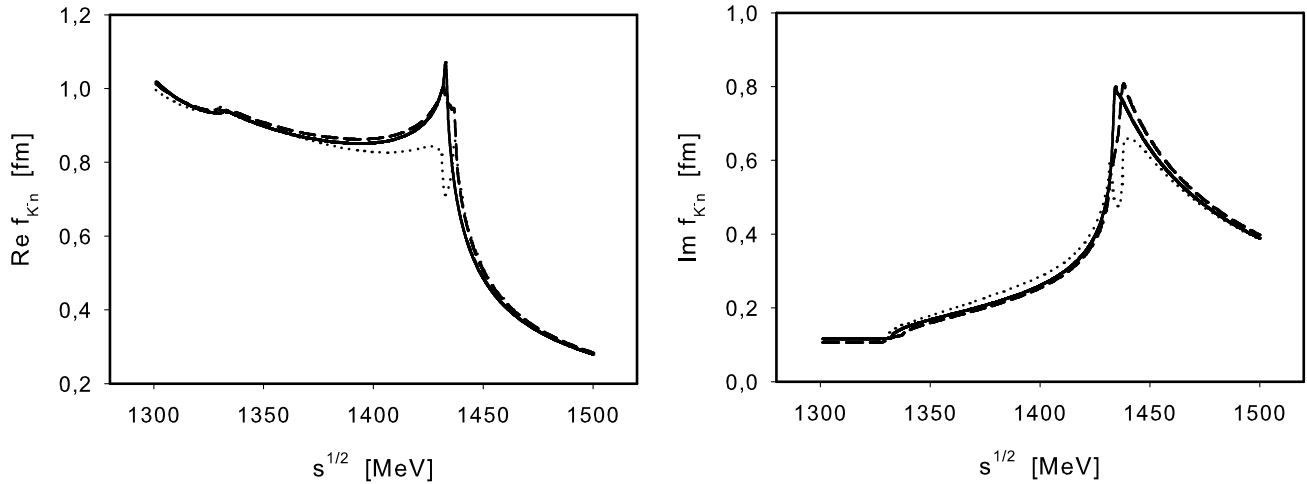}} 
\caption{Comparison of two different isospin schemes with the direct 
calculation of the $K^{-}n$ amplitude. The real and imaginary parts of 
the amplitude are shown in the left and right panels, respectively. 
See text for more details.}
\label{fig:Kn-iso}
\end{figure}

In effect, the $K^{-}n$ amplitude derived by means of isospin relations 
depends on the isospin scheme. We found that it is vital to use the average 
of both, the $K^{-}p$ and $\bar{K}^{0}n$ amplitudes, rather then only 
one of them (normally the $K^{-}p$ one as there are relevant experimental data).
We demonstrate the point in Figure \ref{fig:Kn-iso} where the $K^{-}n$ 
amplitudes obtained by means of two different isospin schemes are compared 
with our direct (six coupled channels) calculation. The dotted lines in 
the figure correspond to the scheme in which the $K^{-}n$ amplitude 
is derived from the elastic $K^{-}p$ amplitude and from the transition 
amplitude of the $K^{-}p \rightarrow \bar{K}^{0}n$ process,
\beq
\langle K^{-}n |f| K^{-}n \rangle = \langle K^{-}p |f| K^{-}p \rangle
     - \langle \bar{K}^{0}n |f| K^{-}p \rangle \; .      
\eeq{eq:Kn-iso} 
One immediately notes the unphysical oscillations in-between the $K^{-}p$ 
and $\bar{K}^{0}n$ thresholds. This can be remedied by taking the average 
of both the elastic $K^{-}p$ and $\bar{K}^{0}n$ amplitudes, {\it i.e.} by replacing 
Eq.(\ref{eq:Kn-iso}) with
\beq
\langle K^{-}n |f| K^{-}n \rangle = (\langle K^{-}p |f| K^{-}p \rangle
       + \langle \bar{K}^{0}n |f| \bar{K}^{0}n \rangle)/2
       - \langle \bar{K}^{0}n |f| K^{-}p \rangle \; .
\eeq{KnKn} 
The resulting $K^{-}n$ amplitude is given by the dashed line in 
Fig.~\ref{fig:Kn-iso}. As this approach does not lead to unphysical 
oscillations it should be prefered in any relevant analysis. This scheme 
is also consistent with the construction of the transition amplitudes between 
the $\bar{K}N$ states of specific isospins when they are decomposed into 
pertinent physical channels. Of course, the dashed lines still exhibit 
cusps at the $K^{-}p$ and $\bar{K}^{0}n$ thresholds and differ slightly 
from the full lines that represent the direct calculation of the $K^{-}n$ 
amplitude (with only one threshold cusp at the physicaly correct 
energy). As expected, the effects related to the isospin violation are 
observed only in the region of the $\bar{K}N$ thresholds and the lines 
practically coincide at energies sufficiently far from the thresholds. 
  
\section{Conclusions}

An effective chirally motivated separable potential was used 
in simultaneous fits of the low energy $K^{-}p$ cross sections, 
the threshold branching ratios and the characteristics of kaonic hydrogen. 
The fits are quite satisfactory except for the 1s level decay width being much 
larger than the experimental value. We have computed the characteristics 
of kaonic hydrogen (the 1s level energy shift and width) exactly and emphasize  
that this approach is vital in view of the expected precision of the coming 
experimental data.

Our results confirm observations by other authors that the coupled channel 
chiral model leads to two poles in the complex energy plane 
that can be related to the 
$\Lambda(1405)$ resonance observed in the $\pi \Sigma$ mass spectrum. However, 
we are not so convinced that both poles affect the physical observables as 
their positions seem to be model dependent and especially the one at higher 
energies may easily drift too far from the real axis. It is also intriguing 
that we were not able to get the position of any of the poles so close to 
the real axis as other authors claim. The disparity can be attributed 
most likely to the inclusion of the $q^{2}$ terms in the chiral Lagrangian 
and partly also to a different formulation of our model, namely to the use 
of effective separable potentials instead of the on-shell scheme employed 
in the inverse T-matrix approach.  

We have also shown that near the $\bar{K}N$ threshold the $K^{-}n$ amplitude 
obtained from multiple channel calculations with physical particle masses 
differs from the one derived by means of isospin relations from 
the transition amplitudes obtained in the $K^{-}p$ sector. 
Despite the underlying chiral Lagrangian adheres to the SU(3) 
symmetry and incorporates fully the isospin symmetry the use of physical masses 
breaks the symmetry. As the thresholds of the $K^{-}p$, $K^{-}n$ and 
$\bar{K}^{0}n$ channels are different one cannot simply relate the $K^{-}n$ amplitude 
to those from the $K^{-}p$ sector. Our results clearly demonstrate this 
and show that both elastic $K^{-}N$ amplitudes exhibit a strong energy dependence 
in the vicinity of $\bar{K}N$ thresholds. It also means that it may not be easy 
and straightforward to relate the $K^{-}$-deuteron scattering length to 
the $K^{-}N$ ones observed in experiments.

We close the paper by expressing a hope that the forthcoming high-precision 
data from the DEAR/SIDDHARTA collaboration and from experiments dedicated 
to the $\Lambda(1405)$ resonance will shred more light 
on kaon-nucleon dynamics and stimulate further theoretical work.  
  
\vspace*{4mm}
{\bf Acknowledgement:} The authors acknowledge the financial support from the 
Grant Agency of the Czech Republic, grant 202/09/1441.
The work of J.~S. was also supported by the Research Program 
{\it Fundamental experiments in the physics of the microworld} No.~6840770040 
of the Ministry of Education, Youth and Sports of the Czech Republic.

\newpage
\centerline{\bf Appendix A}
\renewcommand{\thetable}{A\arabic{table}}
\setcounter{table}{0} 
\bigskip

In the appendices we specify the coefficients $C_{i j}^{\rm (.)}$ 
of Eq.~(\ref{eq:Cij}). First we present the matrices for the channels 
coupled to the $K^{-}p$ ($Q = 0$ and $S = 1$ meson-baryon system), 
the following Appendix B is reserved for the channels coupled to the $K^{-}n$ 
($Q = -1$ and $S = 1$). As the coupling matrices are symmetric, 
$C_{j i}^{\rm (.)} = \, C_{i j}^{\rm (.)}$, we show only the terms above 
the diagonal and cut most tables in two parts to save some space. 
For the later reason we also split the table for the coefficients 
$ C_{i j}^{(EE)}$ in two, so one should sum the respective terms, 
$ C_{i j}^{(EE)} = C_{i j}^{(EE1)} + C_{i j}^{(EE2)}$.

\begin{table}[h]
  \centering{%
    \caption{%
    $K^{-}p$ related channels - the (WT) term coefficients.
        $C_{j i}^{\rm (WT)} = \, C_{i j}^{\rm (WT)}$.}}
  \vspace{0.75cm}

  \begin{tabular}{c|@{\hspace{1em}}cccccccccc}
    & $ \pi^{0} \Lambda$           
    & $ \pi^{0} \Sigma^{0}$        
    & $ \pi^{-} \Sigma^{+} $       
    & $ \pi^{+} \Sigma^{-}$        
    & $ K^{-} p$                   
    & $ \bar{K}^{0} n$             
    & $ \eta \Lambda$              
    & $ \eta \Sigma^{0}$           
    & $ K^{0} \Xi^{0}$             
    & $ K^{+} \Xi^{-}$     \\[2mm] 
  \hline
    $ \pi^{0} \Lambda$ \rule{0em}{1.5em}  
      & 0
      & 0
      & 0
      & 0
      & $\CosXXX$
      & $ - \CosXXX $ 
      & 0
      & 0
      & $ - \CosXXX $
      & $ \CosXXX $                                  \\[2mm]
    $ \pi^{0} \Sigma^{0}$
      &  
      & 0
      & 2
      & 2
      & $\frac{1}{2}$
      & $\frac{1}{2}$ 
      & 0
      & 0
      & $\frac{1}{2}$
      & $\frac{1}{2}$                                \\[2mm]
    $ \pi^{-} \Sigma^{+} $
      &
      & 
      & 2
      & 0
      & 1
      & 0
      & 0
      & 0 
      & 1
      & 0                                            \\[2mm]
    $\pi^{+} \Sigma^{-}$
      &
      &
      &
      & 2
      & 0
      & 1
      & 0
      & 0
      & 0
      & 1                                            \\[2mm]
    $ K^{-} p$ 
      &
      &
      &
      &
      & 2
      & 1
      & $\frac{3}{2}$
      & $ \CosXXX $ 
      & 0
      & 0                                            \\[2mm]
    $ \bar{K}^{0} n$ 
      &
      &
      &
      &
      &
      & 2
      & $\frac{3}{2}$
      & $ - \CosXXX $
      & 0
      & 0                                            \\[2mm]
    $ \eta \Lambda$
      &
      &
      &
      &
      &
      &
      & 0
      & 0 
      & $\frac{3}{2}$
      & $\frac{3}{2}$                                \\[2mm]
    $ \eta \Sigma^{0}$
      &
      &
      &
      &
      &
      &
      &
      & 0 
      & $ - \CosXXX $
      & $ \CosXXX $                                \\[2mm]
    $ K^{0} \Xi^{0}$
      &
      &
      &
      &
      &
      &
      &
      &
      & 2
      & 1                                            \\[2mm]
    $ K^{+} \Xi^{-}$
      &
      &
      &
      &
      &
      &
      &
      &
      &
      & 2
  \end{tabular}
\end{table} 

\begin{table}
  \centering{%
    \caption{%
    $K^{-}p$ related channels - the (mm) term coefficients.
        $C_{j i}^{(mm)} = \, C_{i j}^{(mm)}$.}}
  \vspace{0.75cm}

  \begin{tabular}{c|@{\hspace{1em}}cccccccccc}
    & $ \pi^{0} \Lambda$ \MySepRule                      
    & $ \pi^{0} \Sigma^{0}$                              
    & $ \pi^{-} \Sigma^{+} $                             
    & $ \pi^{+} \Sigma^{-}$                              
    & $ K^{-} p$                              \\[\MySep] 
  \hline
    $ \pi^{0} \Lambda$                \rule{0em}{2.75\MySep}  
      & $\frac{2}{3} (3 \BeNul + \BeDe)$
      & 0
      & 0
      & 0
      & $- \CosXXXthird (\BeDe + 3 \BeFe)$        \\[\MySep]
    $ \pi^{0} \Sigma^{0}$
      &  
      & $2 (\BeNul + \BeDe)$
      & 0
      & 0
      & $\frac{1}{2} (\BeDe - \BeFe)$             \\[\MySep]
    $ \pi^{-} \Sigma^{+} $
      &
      & 
      & $2 (\BeNul + \BeDe)$
      & 0
      & $(\BeDe - \BeFe)$                         \\[\MySep]
    $\pi^{+} \Sigma^{-}$
      &
      &
      &
      & $2 (\BeNul + \BeDe)$
      & 0                                         \\[\MySep]
    $ K^{-} p$ 
      &
      &
      &
      &
      & $2 (\BeNul + \BeDe)$                    \\[.5\MySep]
    \multicolumn{6}{c}{\vspace{1em}}                      \\
    & $ \bar{K}^{0} n$ \MySepRule                       
    & $ \eta \Lambda$                                   
    & $ \eta \Sigma^{0}$                                
    & $ K^{0} \Xi^{0}$                                  
    & $ K^{+} \Xi^{-}$                       \\[\MySep] 
  \hline
    $ \pi^{0} \Lambda$                \rule{0em}{2.75\MySep}
      & $\CosXXXthird (\BeDe + 3 \BeFe)$ 
      & 0 
      & $\frac{2}{3} \BeDe$
      & $\CosXXXthird (\BeDe - 3 \BeFe)$
      & $- \CosXXXthird (\BeDe - 3 \BeFe)$        \\[\MySep]
    $ \pi^{0} \Sigma^{0}$
      & $\frac{1}{2} (\BeDe - \BeFe)$ 
      & $\frac{2}{3} \BeDe$
      & 0
      & $\frac{1}{2} (\BeDe + \BeFe)$
      & $\frac{1}{2} (\BeDe + \BeFe)$             \\[\MySep]
    $ \pi^{-} \Sigma^{+} $
      & 0
      & $\frac{2}{3} \BeDe$
      & $- \frac{2 \sqrt{3}}{3} \BeFe$ 
      & $(\BeDe + \BeFe)$
      & 0                                         \\[\MySep]
    $\pi^{+} \Sigma^{-}$
      & $(\BeDe - \BeFe)$
      & $\frac{2}{3} \BeDe$
      & $\frac{2 \sqrt{3}}{3} \BeFe$
      & 0
      & $(\BeDe + \BeFe)$                         \\[\MySep]
    $ K^{-} p$
      & $(\BeDe + \BeFe)$
      & $\frac{1}{6} (\BeDe + 3\BeFe)$
      & $- \CosXXXthird (\BeDe - \BeFe)$ 
      & 0
      & 0                                         \\[\MySep]
    $ \bar{K}^{0} n$
      & $2 (\BeNul + \BeDe)$
      & $\frac{1}{6} (\BeDe + 3\BeFe)$
      & $\CosXXXthird (\BeDe - \BeFe)$
      & 0
      & 0                                         \\[\MySep]
    $ \eta \Lambda$
      &
      & $2 (\BeNul + \BeDe)$
      & 0 
      & $\frac{1}{6} (\BeDe - 3\BeFe)$
      & $\frac{1}{6} (\BeDe - 3\BeFe)$            \\[\MySep]
    $ \eta \Sigma^{0}$
      &
      &
      & $\frac{2}{3} (3 \BeNul + \BeDe)$ 
      & $\CosXXXthird (\BeDe + \BeFe)$
      & $- \CosXXXthird (\BeDe + \BeFe)$          \\[\MySep]
    $ K^{0} \Xi^{0}$
      &
      &
      &
      & $2 (\BeNul + \BeDe)$
      & $(\BeDe - \BeFe)$                         \\[\MySep]
    $ K^{+} \Xi^{-}$
      & \MinSepRule
      &
      &
      &
      & $2 (\BeNul + \BeDe)$
  \end{tabular}
\end{table}
 
\begin{table}
  \centering{%
  \caption{%
    $K^{-}p$ related channels - the $(\chi {\rm b})$ term coefficients. 
    Only the "lower" part of the table is shown. All coefficients 
    in the "upper" part (terms above the diagonal in columns from 
    $\pi^{0} \Lambda$ to $ K^{-} p$) are equal to zero. 
    $C_{j i}^{(\chi {\rm b})} = \, C_{i j}^{(\chi {\rm b})}$.
    }}
  \vspace{0.75cm}

  \begin{tabular}{c|@{\hspace{1em}}cccccccccc}
    & $ \bar{K}^{0} n$ \MySepRule                       
    & $ \eta \Lambda$                                   
    & $ \eta \Sigma^{0}$                                
    & $ K^{0} \Xi^{0}$                                  
    & $ K^{+} \Xi^{-}$                       \\[\MySep] 
  \hline
    $ \pi^{0} \Lambda$                \rule{0em}{2.75\MySep}
      & 0 
      & $\frac{4}{9}(\BeDe +3\BeFe)$
      & -$\frac{4\sqrt{3}}{9}(\BeDe -\BeFe)$
      & 0
      & 0                                         \\[\MySep]
    $ \pi^{0} \Sigma^{0}$
      & 0
      & -$\frac{8}{9}\BeDe$
      & 0
      & 0
      & 0                                         \\[\MySep]
    $ \pi^{-} \Sigma^{+} $
      & 0
      & $-\frac{8}{9}\BeDe$
      & $\frac{8\sqrt{3}}{9}\BeFe$
      & 0
      & 0                                         \\[\MySep]
    $\pi^{+} \Sigma^{-}$
      & 0
      & $-\frac{8}{9}\BeDe$
      & $-\frac{8\sqrt{3}}{9}\BeFe$
      & 0
      & 0                                         \\[\MySep]
    $ K^{-} p$
      & 0
      & $\frac{4}{9}(\BeDe +3\BeFe)$
      & $-\frac{4\sqrt{3}}{9}(\BeDe -\BeFe)$ 
      & 0
      & 0                                         \\[\MySep]
    $ \bar{K}^{0} n$
      & 0
      & $\frac{4}{9}(\BeDe +3\BeFe)$
      & $\frac{4}{9}(\BeDe -\BeFe)$
      & 0
      & 0                                         \\[\MySep]
    $ \eta \Lambda$
      & 
      & $\frac{16}{9}\BeDe$
      & 0 
      & $\frac{4}{9}(\BeDe -3\BeFe)$
      & $\frac{4}{9}(\BeDe -3\BeFe)$              \\[\MySep]
    $ \eta \Sigma^{0}$
      &
      &
      & $-\frac{16}{9}\BeDe$ 
      & $\frac{4\sqrt{3}}{9}(\BeDe +\BeFe)$
      & $-\frac{4\sqrt{3}}{9}(\BeDe +\BeFe)$      \\[\MySep]
    $ K^{0} \Xi^{0}$
      &
      &
      &
      & 0
      & 0                                         \\[\MySep]
    $ K^{+} \Xi^{-}$
      & \MinSepRule
      &
      &
      &
      & 0
  \end{tabular}
\end{table}

\begin{table}
  \centering{%
  \caption{%
      $K^{-}p$ related channels - the (EE1) term coefficients.
      $C_{j i}^{(EE1)} = C_{i j}^{(EE1)}$.}}
  \vspace{0.75cm}

  \begin{tabular}{c|@{\hspace{1em}}cccccccccc}
    & $ \pi^{0} \Lambda$ \MySepRule                      
    & $ \pi^{0} \Sigma^{0}$                              
    & $ \pi^{-} \Sigma^{+} $                             
    & $ \pi^{+} \Sigma^{-}$                              
    & $ K^{-} p$                              \\[\MySep] 
  \hline
    $ \pi^{0} \Lambda$                \rule{0em}{2.75\MySep}  
      & $- \frac{2}{3} (3\DeNul + \DeDe)$
      & 0
      & 0
      & 0
      & $\CosXXXthird (\DeDe + 3 \DeFe)$         \\[\MySep]
    $ \pi^{0} \Sigma^{0}$
      &  
      & $- 2 (\DeNul + \DeDe)$
      & 0
      & 0
      & $- \frac{1}{2} (\DeDe - \DeFe)$           \\[\MySep]
    $ \pi^{-} \Sigma^{+} $
      &
      & 
      & $- 2 (\DeNul + \DeDe)$
      & 0
      & $- (\DeDe - \DeFe)$                       \\[\MySep]
    $\pi^{+} \Sigma^{-}$
      &
      &
      &
      & $- 2 (\DeNul + \DeDe)$
      & 0                                         \\[\MySep]
    $ K^{-} p$ 
      &
      &
      &
      &
      & $- 2 (\DeNul + \DeDe)$                  \\[.5\MySep]
    \multicolumn{6}{c}{\vspace{1em}}                      \\
    & $ \bar{K}^{0} n$ \MySepRule                       
    & $ \eta \Lambda$                                   
    & $ \eta \Sigma^{0}$                                
    & $ K^{0} \Xi^{0}$                                  
    & $ K^{+} \Xi^{-}$                       \\[\MySep] 
  \hline
    $ \pi^{0} \Lambda$                \rule{0em}{2.75\MySep}
      & $- \CosXXXthird (\DeDe + 3 \DeFe)$ 
      & 0 
      & $- \frac{2}{3} \DeDe$
      & $- \CosXXXthird (\DeDe - 3 \DeFe)$
      & $\CosXXXthird (\DeDe - 3 \DeFe)$          \\[\MySep]
    $ \pi^{0} \Sigma^{0}$
      & $- \frac{1}{2} (\DeDe - \DeFe)$ 
      & $- \frac{2}{3} \DeDe$
      & 0
      & $- \frac{1}{2} (\DeDe + \DeFe)$
      & $- \frac{1}{2} (\DeDe + \DeFe)$           \\[\MySep]
    $ \pi^{-} \Sigma^{+} $
      & 0
      & $- \frac{2}{3} \DeDe$
      & $\frac{2 \sqrt{3}}{3} \DeFe$ 
      & $- (\DeDe + \DeFe)$
      & 0                                         \\[\MySep]
    $\pi^{+} \Sigma^{-}$
      & $- (\DeDe - \DeFe)$
      & $- \frac{2}{3} \DeDe$
      & $- \frac{2 \sqrt{3}}{3} \DeFe$
      & 0
      & $- (\DeDe + \DeFe)$                       \\[\MySep]
    $ K^{-} p$
      & $- (\DeDe + \DeFe)$
      & $- \frac{1}{6} (\DeDe + 3\DeFe)$
      & $\CosXXXthird (\DeDe - \DeFe)$ 
      & 0
      & 0                                         \\[\MySep]
    $ \bar{K}^{0} n$
      & $- 2 (\DeNul + \DeDe)$
      & $- \frac{1}{6} (\DeDe + 3\DeFe)$
      & $- \CosXXXthird (\DeDe - \DeFe)$
      & 0
      & 0                                         \\[\MySep]
    $ \eta \Lambda$
      &
      & $- 2 (\DeNul + \DeDe)$
      & 0 
      & $- \frac{1}{6} (\DeDe - 3\DeFe)$
      & $- \frac{1}{6} (\DeDe - 3\DeFe)$          \\[\MySep]
    $ \eta \Sigma^{0}$
      &
      &
      & $- \frac{2}{3} (3 \DeNul + \DeDe)$ 
      & $- \CosXXXthird (\DeDe + \DeFe)$
      & $\CosXXXthird (\DeDe + \DeFe)$            \\[\MySep]
    $ K^{0} \Xi^{0}$
      &
      &
      &
      & $- 2 (\DeNul + \DeDe)$
      & $- (\DeDe - \DeFe)$                       \\[\MySep]
    $ K^{+} \Xi^{-}$
      & \MinSepRule
      &
      &
      &
      & $- 2 (\DeNul + \DeDe)$
  \end{tabular}
\end{table}
 
\begin{table}
  \centering{%
  \caption{%
      $K^{-}p$ related channels - the (EE2) term coefficients.
      $C_{j i}^{(EE2)} = C_{i j}^{(EE2)}$.}}
  \vspace{0.75cm}

  \begin{tabular}{c|@{\hspace{1em}}cccccccccc}
    & $ \pi^{0} \Lambda$ \MySepRule                      
    & $ \pi^{0} \Sigma^{0}$                              
    & $ \pi^{-} \Sigma^{+} $                             
    & $ \pi^{+} \Sigma^{-}$                              
    & $ K^{-} p$                              \\[\MySep] 
  \hline
    $ \pi^{0} \Lambda$                \rule{0em}{2.75\MySep}  
      & $-\frac{1}{3}\DeII$
      & 0
      & 0
      & 0
      & $-\CosXXXthird \DeII$                 \\[\MySep]
    $ \pi^{0} \Sigma^{0}$
      &  
      & $-2\DeI -\DeII$
      & $-\DeI -\DeII$
      & $-\DeI -\DeII$
      & $-\DeI -\frac{1}{2}\DeII$             \\[\MySep]
    $ \pi^{-} \Sigma^{+} $
      &
      & 
      & $-\DeI$
      & $-2\DeI -2\DeII$
      & $-\DeI$                               \\[\MySep]
    $\pi^{+} \Sigma^{-}$
      &
      &
      &
      & $-\DeI$
      & $-\DeI -\DeII$                                         \\[\MySep]
    $ K^{-} p$ 
      &
      &
      &
      &
      & $-\DeI$                     \\[.5\MySep]
    \multicolumn{6}{c}{\vspace{1em}}                      \\
    & $ \bar{K}^{0} n$ \MySepRule                       
    & $ \eta \Lambda$                                   
    & $ \eta \Sigma^{0}$                                
    & $ K^{0} \Xi^{0}$                                  
    & $ K^{+} \Xi^{-}$                       \\[\MySep] 
  \hline
    $ \pi^{0} \Lambda$                \rule{0em}{2.75\MySep}
      & $\CosXXXthird \DeII$ 
      & 0 
      & $-\DeI -\frac{1}{3}\DeII$
      & $\CosXXXthird \DeII$
      & $-\CosXXXthird \DeII$             \\[\MySep]
    $ \pi^{0} \Sigma^{0}$
      & $-\DeI -\frac{1}{2}\DeII$ 
      & $-\DeI -\frac{1}{3}\DeII$
      & 0
      & $-\DeI -\frac{1}{2}\DeII$
      & $-\DeI -\frac{1}{2}\DeII$           \\[\MySep]
    $ \pi^{-} \Sigma^{+} $
      & $-\DeI -\DeII$
      & $-\DeI -\frac{1}{3}\DeII$
      & 0 
      & $-\DeI$
      & $-\DeI -\DeII$                           \\[\MySep]
    $\pi^{+} \Sigma^{-}$
      & $-\DeI$
      & $-\DeI -\frac{1}{3}\DeII$
      & 0
      & $-\DeI -\DeII$
      & $-\DeI$                       \\[\MySep]
    $ K^{-} p$
      & $-\DeI$
      & $-\DeI -\frac{5}{6}\DeII$
      & $-\CosXXXthird \DeII$ 
      & $-\DeI -\DeII$
      & $-2(\DeI +\DeII)$                         \\[\MySep]
    $ \bar{K}^{0} n$
      & $-\DeI$
      & $-\DeI -\frac{5}{6}\DeII$
      & $\CosXXXthird \DeII$
      & $-2(\DeI +\DeII)$
      & $-\DeI -\DeII$                            \\[\MySep]
    $ \eta \Lambda$
      &
      & $-2\DeI -\DeII$
      & 0 
      & $-\DeI -\frac{5}{6}\DeII$
      & $-\DeI -\frac{5}{6}\DeII$                 \\[\MySep]
    $ \eta \Sigma^{0}$
      &
      &
      & $-\frac{1}{3}\DeII$ 
      & $\CosXXXthird \DeII$
      & $-\CosXXXthird \DeII$                     \\[\MySep]
    $ K^{0} \Xi^{0}$
      &
      &
      &
      & $-\DeI$
      & $-\DeI$                                   \\[\MySep]
    $ K^{+} \Xi^{-}$
      & \MinSepRule
      &
      &
      &
      & $-\DeI$
  \end{tabular}
\end{table}

\begin{table}
  \centering{%
  \caption{%
    $K^{-}p$ related channels - the direct (s) term coefficients.
      $C_{j i}^{(s)} = C_{i j}^{(s)}$.}}
  \vspace{0.75cm}

  \begin{tabular}{c|cccccccccc}
    & $ \pi^{0} \Lambda$ \MySepRule                      
    & $ \pi^{0} \Sigma^{0}$                              
    & $ \pi^{-} \Sigma^{+} $                             
    & $ \pi^{+} \Sigma^{-}$                              
    & $ K^{-} p$                              \\[\MySep] 
  \hline
    $ \pi^{0} \Lambda$                \rule{0em}{2.75\MySep}  
      & $-\frac{1}{3}\DxD$
      & 0
      & $-\frac{\sqrt{3}}{3}\DxF$
      & $\frac{\sqrt{3}}{3}\DxF$
      & $\frac{\sqrt{3}}{6}(\DxD \!-\! \DxF)$           \\[\MySep]
    $ \pi^{0} \Sigma^{0}$
      &  
      & $\frac{1}{3}\DxD$
      & $\frac{1}{3}\DxD$
      & $\frac{1}{3}\DxD$
      & $-\frac{1}{6}\DxD \!-\! \frac{1}{2}\DxF$       \\[\MySep]
    $ \pi^{-} \Sigma^{+} $
      &
      & 
      & $\frac{1}{3}\DxD +\FxF$
      & $\frac{1}{3}\DxD -\FxF$
      & $-\frac{1}{6}\DxD \!-\! \DxF \!+\! \frac{1}{2}\FxF$  \\[\MySep]
    $\pi^{+} \Sigma^{-}$
      &
      &
      &
      & $\frac{1}{3}\DxD \!+\! \FxF$
      & $-\frac{1}{6}\DxD \!-\! \frac{1}{2}\FxF$       \\[\MySep]
    $ K^{-} p$ 
      &
      &
      &
      &
      & $\frac{1}{3}\DxD \!+\! \FxF$                 \\[.5\MySep]
    \multicolumn{6}{c}{\vspace{1em}}                      \\
    & $ \bar{K}^{0} n$ \MySepRule                       
    & $ \eta \Lambda$                                   
    & $ \eta \Sigma^{0}$                                
    & $ K^{0} \Xi^{0}$                                  
    & $ K^{+} \Xi^{-}$                       \\[\MySep] 
  \hline
    $ \pi^{0} \Lambda$                \rule{0em}{2.75\MySep}
      & $-\frac{\sqrt{3}}{6}(\DxD \!-\! \DxF)$
      & 0 
      & $\frac{1}{3}\DxD$
      & $-\frac{\sqrt{3}}{6}(\DxD \!+\! \DxF)$
      & $\frac{\sqrt{3}}{6}(\DxD \!+\! \DxF)$          \\[\MySep]
    $ \pi^{0} \Sigma^{0}$
      & $-\frac{1}{6}(\DxD \!+\! 3\DxF)$
      & $-\frac{1}{3}\DxD$
      & 0
      & $-\frac{1}{6}(\DxD \!-\! 3\DxF)$
      & $-\frac{1}{6}(\DxD \!-\! 3\DxF)$         \\[\MySep]
    $ \pi^{-} \Sigma^{+} $
      & $-\frac{1}{6}(\DxD \!+\! 3\FxF)$
      & $-\frac{1}{3}\DxD$
      & $-\frac{\sqrt{3}}{3}\DxF$
      & $-\frac{1}{6}\DxD \!+\! \frac{1}{2}\FxF \!+\! \DxF$
      & $-\frac{1}{6}(\DxD \!+\! 3\FxF)$                  \\[\MySep]
    $\pi^{+} \Sigma^{-}$
      & $-\frac{1}{6}\DxD \!+\! \frac{1}{2}\FxF \!-\! \DxF$
      & $-\frac{1}{3}\DxD$
      & $\frac{\sqrt{3}}{3}\DxF$
      & $-\frac{1}{6}(\DxD \!+\! 3\FxF)$
      & $-\frac{1}{6}\DxD \!+\! \frac{1}{2}\FxF \!+\! \DxF$        \\[\MySep]
    $ K^{-} p$
      & $-\frac{1}{6}\DxD \!+\! \frac{1}{2}\FxF \!+\! \DxF$
      & $\frac{1}{6}(\DxD \!+\! 3\DxF)$
      & $\frac{\sqrt{3}}{6}(\DxD \!-\! \DxF)$
      & $-\frac{1}{6}(\DxD \!+\! 3\FxF)$
      & $\frac{1}{3}(\DxD \!-\! 3\FxF)$                     \\[\MySep]
    $ \bar{K}^{0} n$
      & $\frac{1}{3}(\DxD \!+\! 3\FxF)$
      & $\frac{1}{6}(\DxD \!+\! 3\DxF)$
      & $-\frac{\sqrt{3}}{6}(\DxD \!-\! \DxF)$
      & $\frac{1}{3}(\DxD \!-\! 3\FxF)$
      & $-\frac{1}{6}(\DxD \!+\! 3\FxF)$                            \\[\MySep]
    $ \eta \Lambda$
      &
      & $\frac{1}{3}\DxD$
      & 0 
      & $\frac{1}{6}(\DxD \!-\! 3\DxF)$
      & $\frac{1}{6}(\DxD \!-\! 3\DxF)$                 \\[\MySep]
    $ \eta \Sigma^{0}$
      &
      &
      & $\frac{1}{3}\DxD$ 
      & $-\frac{\sqrt{3}}{6}(\DxD \!+\! \DxF)$
      & $\frac{\sqrt{3}}{6}(\DxD \!+\! \DxF)$               \\[\MySep]
    $ K^{0} \Xi^{0}$
      &
      &
      &
      & $\frac{1}{3}(\DxD \!+\! 3\FxF)$
      & $-\frac{1}{6}\DxD \!+\! \frac{1}{2}\FxF \!-\! \DxF$                       \\[\MySep]
    $ K^{+} \Xi^{-}$
      & \MinSepRule
      &
      &
      &
      & $\frac{1}{3}(\DxD \!+\! 3\FxF)$
  \end{tabular}
\end{table}

\begin{table}
  \centering{%
  \caption{%
      $K^{-}p$ related channels - the crossed (u) term coefficients. 
      $C_{j i}^{(u)} = C_{i j}^{(u)}$. 
      To shorten the length of some coefficients we denote 
      ${\cal U}_{+++} = \frac{\sqrt{3}}{12}\DxD + \frac{\sqrt{3}}{4}\FxF 
             + \frac{\sqrt{3}}{3}\DxF$ 
      and use ${\cal U}_{abc}$ with $a$, $b$ and $c$ marking the signs of 
      the terms with $\DxD$, $\FxF$ and $\DxF$, respectively.   
      }}
  \vspace{0.75cm}

  \begin{tabular}{c|@{\hspace{1em}}cccccccccc}
    & $ \pi^{0} \Lambda$ \MySepRule                      
    & $ \pi^{0} \Sigma^{0}$                              
    & $ \pi^{-} \Sigma^{+} $                             
    & $ \pi^{+} \Sigma^{-}$                              
    & $ K^{-} p$                              \\[\MySep] 
  \hline
    $ \pi^{0} \Lambda$                \rule{0em}{2.75\MySep}  
      & $\frac{1}{3}\DxD$
      & 0
      & $\frac{\sqrt{3}}{3}\DxF$
      & $-\frac{\sqrt{3}}{3}\DxF$
      & ${\cal U}_{---}$                          \\[\MySep]
    $ \pi^{0} \Sigma^{0}$
      &  
      & $\frac{1}{3}\DxD$
      & $-\FxF$
      & $-\FxF$
      & $\frac{1}{4}(\DxD - \FxF)$       \\[\MySep]
    $ \pi^{-} \Sigma^{+} $
      &
      & 
      & 0
      & $\frac{1}{3}\DxD -\FxF$
      & 0                                         \\[\MySep]
    $\pi^{+} \Sigma^{-}$
      &
      &
      &
      & 0
      & $\frac{1}{2}(\DxD - \FxF)$       \\[\MySep]
    $ K^{-} p$ 
      &
      &
      &
      &
      & 0                                  \\[.5\MySep]
    \multicolumn{6}{c}{\vspace{1em}}                      \\
    & $ \bar{K}^{0} n$ \MySepRule                       
    & $ \eta \Lambda$                                   
    & $ \eta \Sigma^{0}$                                
    & $ K^{0} \Xi^{0}$                                  
    & $ K^{+} \Xi^{-}$                       \\[\MySep] 
  \hline
    $ \pi^{0} \Lambda$                \rule{0em}{2.75\MySep}
      & ${\cal U}_{+++}$
      & 0 
      & $-\frac{1}{3}\DxD$
      & ${\cal U}_{++-}$
      & ${\cal U}_{--+}$          \\[\MySep]
    $ \pi^{0} \Sigma^{0}$
      & $\frac{1}{4}(\DxD -\FxF)$
      & $\frac{1}{3}\DxD$
      & 0
      & $\frac{1}{4}(\DxD - \FxF)$
      & $\frac{1}{4}(\DxD - \FxF)$         \\[\MySep]
    $ \pi^{-} \Sigma^{+} $
      & $\frac{1}{2}(\DxD - \FxF)$
      & $-\frac{1}{3}\DxD$
      & $-\frac{\sqrt{3}}{3}\DxF$
      & 0
      & $\frac{1}{2}(\DxD - \FxF)$                \\[\MySep]
    $\pi^{+} \Sigma^{-}$
      & 0
      & $\frac{1}{3}\DxD$
      & $\frac{\sqrt{3}}{3}\DxF$
      & $\frac{1}{2}(\DxD -\FxF)$
      & 0                                         \\[\MySep]
    $ K^{-} p$
      & 0
      & $\frac{1}{12}(\DxD - 9\FxF)$
      & ${\cal U}_{--+}$
      & $\frac{1}{2}(\DxD - \FxF)$
      & $\frac{1}{3}(\DxD - 3\FxF)$                     \\[\MySep]
    $ \bar{K}^{0} n$
      & 0
      & $\frac{1}{12}(\DxD - 9\FxF)$
      & ${\cal U}_{++-}$
      & $\frac{1}{3}(\DxD -3\FxF)$
      & $\frac{1}{2}(\DxD - \FxF)$                            \\[\MySep]
    $ \eta \Lambda$
      &
      & $\frac{1}{3}\DxD$
      & 0 
      & $\frac{1}{12}(\DxD -9\FxF)$
      & $\frac{1}{12}(\DxD -9\FxF)$                 \\[\MySep]
    $ \eta \Sigma^{0}$
      &
      &
      & $\frac{1}{3}\DxD$ 
      & ${\cal U}_{+++}$
      & ${\cal U}_{---}$               \\[\MySep]
    $ K^{0} \Xi^{0}$
      &
      &
      &
      & 0
      & 0                                        \\[\MySep]
    $ K^{+} \Xi^{-}$
      & \MinSepRule
      &
      &
      &
      & 0
  \end{tabular}
\end{table}

\newpage
\centerline{\bf Appendix B}
\renewcommand{\thetable}{B\arabic{table}}
\setcounter{table}{0} 
\bigskip

Here we present the coefficients $C_{i j}^{\rm (.)}$ 
of Eq.~(\ref{eq:Cij}) for the six channels coupled to the $K^{-}n$ system 
($Q = -1$ and $S = 1$). The couplings have exactly the same 
structure as in the $K^{-}p$ case and are symmetric too, 
$C_{j i}^{\rm (.)} = \, C_{i j}^{\rm (.)}$.  
\bigskip

\begin{table}[h]
  \centering
  \caption{
    $K^{-}n$ related channels - the (WT) term coefficients.
    }
  \vspace{0.75cm}

  \begin{tabular}{c|cccccc}            
    & $\pi^{-} \Lambda$       
    & $\pi^{-} \Sigma^{0}$    
    & $\pi^{0} \Sigma^{-}$    
    & $K^{-}   n$             
    & $\eta    \Sigma^{-}$    
    & $K^{0}   \Xi^{-}$       
    \\[2mm] \hline
    $\pi^{-} \Lambda$                 \rule{0em}{2.75\MySep}
      & 0
      & 0
      & 0
      & $\sqrt{3/2}$
      & 0
      & $\sqrt{3/2}$                              \\[\MySep]
    $\pi^{-} \Sigma^{0}$
      & 
      & 0
      & -2
      & $-\sqrt{1/2}$
      & 0
      & $\sqrt{1/2}$                              \\[\MySep]
    $\pi^{0} \Sigma^{-}$
      &
      & 
      & 0
      & $\sqrt{1/2}$
      & 0
      & $-\sqrt{1/2}$                             \\[\MySep]
    $K^{-} n$
      &
      &
      &
      & 1
      & $\sqrt{3/2}$
      & 0                                         \\[\MySep]
    $\eta \Sigma^{-}$ 
      &
      &
      &
      &
      & 0
      & $\sqrt{3/2}$                              \\[\MySep]
    $K^{0} \Xi^{-}$ 
      &
      &
      &
      &
      &
      & 1
  \end{tabular}

\end{table} 

\begin{table}[h]
  \centering
  \caption{%
    $K^{-}n$ related channels - the (mm) term coefficients.
    }
  \vspace{0.75cm}

  \begin{tabular}{c|cccccc}
    & $ \pi^{-} \Lambda$              
    & $ \pi^{-} \Sigma^{0}$           
    & $ \pi^{0} \Sigma^{-}$           
    & $ K^{-} n$                      
    & $ \eta \Sigma^{-}$              
    & $ K^{0} \Xi^{-}$ \\[\MySep]     
  \hline
    $ \pi^{-} \Lambda$                \rule{0em}{2.75\MySep}  
      & $\frac{2}{3} (3 \BeNul + \BeDe)$
      & 0
      & 0
      & $- \RootOfSixth (\BeDe + 3 \BeFe)$
      & $\frac{2}{3} \BeDe$
      & $- \RootOfSixth (\BeDe - 3 \BeFe)$        \\[\MySep]
    $ \pi^{-} \Sigma^{0}$
      &  
      & $2 (\BeNul + \BeDe)$
      & 0
      & $- \CosXLV (\BeDe - \BeFe)$
      & $\frac{2 \sqrt{3}}{3} \BeFe$
      & $\CosXLV (\BeDe + \BeFe)$                 \\[\MySep]
    $ \pi^{0} \Sigma^{-}$
      &
      & 
      & $2 (\BeNul + \BeDe)$
      & $\CosXLV (\BeDe - \BeFe)$
      & $- \frac{2 \sqrt{3}}{3} \BeFe$
      & $- \CosXLV (\BeDe + \BeFe)$               \\[\MySep]
    $ K^{-} n$
      &
      &
      &
      & $(2 \BeNul + \BeDe - \BeFe)$
      & $- \RootOfSixth (\BeDe - \BeFe)$
      & 0                                         \\[\MySep]
    $ \eta \Sigma^{-}$ 
      &
      &
      &
      &
      & $\frac{2}{3} (3 \BeNul + \BeDe)$
      & $- \RootOfSixth (\BeDe + \BeFe)$          \\[\MySep]
    $ K^{0} \Xi^{-}$ 
      &
      &
      &
      &
      &
      & $(2 \BeNul + \BeDe + \BeFe)$
  \end{tabular}

\end{table} 

\begin{table}
  \centering
  \caption{%
    $K^{-}n$ related channels - the $(\chi {\rm b})$ term coefficients.
    }
  \vspace{0.75cm}

  \begin{tabular}{c|cccccc}
    \MySepRule
      & $ \pi^{-} \Lambda$                               
      & $ \pi^{-} \Sigma^{0}$                            
      & $ \pi^{0} \Sigma^{-}$                            
      & $ K^{-} n$                                       
      & $ \eta \Sigma^{-}$                               
      & $ K^{0} \Xi^{-}$                      \\[\MySep] 
    \hline
    $ \pi^{-} \Lambda$                \rule{0em}{2.75\MySep}  
      & 0
      & 0
      & 0
      & 0
      & $- \frac{8}{9} \BeDe$
      & 0                                         \\[\MySep]
    $ \pi^{-} \Sigma^{0}$
      &  
      & 0
      & 0
      & 0
      & $- \frac{8 \sqrt{3}}{9} \BeFe$
      & 0                                         \\[\MySep]
    $ \pi^{0} \Sigma^{-}$
      &
      & 
      & 0
      & 0
      & $\frac{8 \sqrt{3}}{9} \BeFe$
      & 0                                         \\[\MySep]
    $ K^{-} n$
      &
      &
      &
      & 0
      & $- \frac{4 \sqrt{6}}{9} (\BeDe - \BeFe)$
      & 0                                         \\[\MySep]
    $ \eta \Sigma^{-}$ 
      &
      &
      &
      &
      & $- \frac{16}{9} \BeDe$
      & $- \frac{4 \sqrt{6}}{9} (\BeDe + \BeFe)$  \\[\MySep]
    $ K^{0} \Xi^{-}$ 
      & \MyHBox
      & \MyHBox
      & \MyHBox
      & \MyHBox
      & \MinSepRule
      & 0
  \end{tabular}
\end{table}

\begin{table}
  \centering
  \caption{%
      $K^{-}n$ related channels - the (EE1) term coefficients.
      $C_{j i}^{(EE1)} \!= C_{i j}^{(EE1)}$.}
  \vspace{0.75cm}
  \begin{tabular}%
      {@{}c|c@{\hspace{.8em}}c@{\hspace{.8em}}cccc@{}}
    & $ \pi^{-} \Lambda$                                 
    & $ \pi^{-} \Sigma^{0}$                              
    & $ \pi^{0} \Sigma^{-}$                              
    & $ K^{-} n$                                         
    & $ \eta \Sigma^{-}$                                 
    & $ K^{0} \Xi^{-}$                        \\[\MySep] 
  \hline
    $ \pi^{-} \Lambda$                \rule{0em}{2.75\MySep}  
      & $- \frac{2}{3} \DeDe$
      & 0
      & 0
      & $\RootOfSixth (\DeDe + 3\DeFe)$
      & $- \frac{2}{3} \DeDe$
      & $\RootOfSixth (\DeDe - 3\DeFe)$      \\[\MySep]
    $ \pi^{-} \Sigma^{0}$
      & \MyHBox 
      & $- 2 \DeDe$
      & 0
      & $\CosXLV (\DeDe - \DeFe)$
      & $- \frac{2 \sqrt{3}}{3} \DeFe$
      & $- \CosXLV (\DeDe + \DeFe)$               \\[\MySep]
    $ \pi^{0} \Sigma^{-}$
      & \MyHBox
      & \MyHBox
      & $- 2 \DeDe$
      & $- \CosXLV (\DeDe - \DeFe)$
      & $\frac{2 \sqrt{3}}{3} \DeFe$
      & $\CosXLV (\DeDe + \DeFe)$                 \\[\MySep]
    $ K^{-} n$
      & \MyHBox
      & \MyHBox
      & \MyHBox
      & $- (\DeDe - \DeFe)$
      & $\RootOfSixth (\DeDe - \DeFe)$
      & 0                                         \\[\MySep]
    $ \eta \Sigma^{-}$ 
      &
      &
      &
      &
      & $- \frac{2}{3} \DeDe$
      & $\RootOfSixth (\DeDe + \DeFe)$            \\[\MySep]
    $ K^{0} \Xi^{-}$ 
      &
      &
      &
      &
      &
      & $- (\DeDe + \DeFe)$
  \end{tabular}
\end{table}

\begin{table}
  \centering
  \caption{%
    $K^{-}n$ related channels - the (EE2) term coefficients.
      $C_{j i}^{(EE2)} \!= C_{i j}^{(EE2)}$.}
  \vspace{0.75cm}
  \begin{tabular}%
      {@{}c|c@{\hspace{.8em}}c@{\hspace{.8em}}cccc@{}}
    & $ \pi^{-} \Lambda$                                 
    & $ \pi^{-} \Sigma^{0}$                              
    & $ \pi^{0} \Sigma^{-}$                              
    & $ K^{-} n$                                         
    & $ \eta \Sigma^{-}$                                 
    & $ K^{0} \Xi^{-}$                        \\[\MySep] 
  \hline
    $ \pi^{-} \Lambda$                \rule{0em}{2.75\MySep}  
      & $- 2\DeNul - \frac{1}{3} \DeII$
      & 0
      & 0
      & $- \RootOfSixth \DeII$
      & $- \frac{1}{3} (3 \DeI + \DeII)$
      & $- \RootOfSixth \DeII$                    \\[\MySep]
    $ \pi^{-} \Sigma^{0}$
      &  
      & $- 2\DeNul + \DeII$
      & $- (\DeI + \DeII)$
      & $- \CosXLV \DeII$
      & 0
      & $\CosXLV \DeII$                           \\[\MySep]
    $ \pi^{0} \Sigma^{-}$
      &
      & 
      & $- 2\DeNul + \DeII$
      & $\CosXLV \DeII$
      & 0
      & $- \CosXLV \DeII$                         \\[\MySep]
    $ K^{-} n$
      &
      &
      &
      & $- 2\DeNul$
      & $- \RootOfSixth \DeII$
      & $- (\DeI + \DeII)$                        \\[\MySep]
    $ \eta \Sigma^{-}$ 
      &
      &
      &
      &
      & $- 2\DeNul - \frac{1}{3} \DeII$
      & $- \RootOfSixth \DeII$                    \\[\MySep]
    $ K^{0} \Xi^{-}$ 
      & \MyHBox
      & \MyHBox
      & \MyHBox
      & \MyHBox
      & \MinSepRule
      & $- 2\DeNul$
  \end{tabular}
\end{table}

\begin{table}
  \centering
  \caption{%
    $K^{-}n$ related channels - the direct (s) term coefficients.
      $C_{j i}^{(s)} \!= C_{i j}^{(s)}$.}
  \vspace{0.75cm}

  \begin{tabular}%
      {c|cccccc}
    \MySepRule
      & $ \pi^{-} \Lambda$                               
      & $ \pi^{-} \Sigma^{0}$                            
      & $ \pi^{0} \Sigma^{-}$                            
      & $ K^{-} n$                                       
      & $ \eta \Sigma^{-}$                               
      & $ K^{0} \Xi^{-}$                      \\[\MySep] 
    \hline
    $ \pi^{-} \Lambda$                \rule{0em}{2.75\MySep}  
      & $\frac{1}{3} \DxD$
      & $\TanXXX \DxF$
      & $- \TanXXX \DxF$
      & $\RootOfSixth (\DxD - \DxF)$
      & $\frac{1}{3} \DxD$
      & $\RootOfSixth (\DxD + \DxF)$              \\[\MySep]
    $ \pi^{-} \Sigma^{0}$
      &  
      & $\FxF$
      & $- \FxF$
      & $\CosXLV (\DxF - \FxF)$
      & $\TanXXX \DxF$
      & $\CosXLV (\DxF + \FxF)$                   \\[\MySep]
    $ \pi^{0} \Sigma^{-}$
      &
      & 
      & $\FxF$
      & $- \CosXLV (\DxF - \FxF)$
      & $- \TanXXX \DxF$
      & $- \CosXLV (\DxF + \FxF)$                 \\[\MySep]
    $ K^{-} n$
      &
      &
      &
      & $\frac{1}{2} (\DxD - 2 \DxF + \FxF)$
      & $\RootOfSixth (\DxD - \DxF)$
      & $\frac{1}{2} (\DxD - \FxF)$               \\[\MySep]
    $ \eta \Sigma^{-}$ 
      &
      &
      &
      &
      & $\frac{1}{3} \DxD$
      & $\RootOfSixth (\DxD + \DxF)$              \\[\MySep]
    $ K^{0} \Xi^{-}$ 
      &
      &
      &
      &
      & \MinSepRule
      & $\frac{1}{2} (\DxD + 2 \DxF + \FxF)$
  \end{tabular}
\end{table}
 
\begin{table}
  \centering
  \caption{%
    $K^{-}n$ related channels - the crossed (u) term coefficients.
      $C_{j i}^{(u)} \!= C_{i j}^{(u)}$.}
  \vspace{0.75cm}

  \begin{tabular}%
      {@{}c@{\hspace{.325em}}|%
        cccc@{\hspace{-.225em}}c@{\hspace{.5em}}c@{}}
    \MySepRule
      & $ \pi^{-} \Lambda$                               
      & $ \pi^{-} \Sigma^{0}$                            
      & $ \pi^{0} \Sigma^{-}$                            
      & $ K^{-} n$                                       
      & $ \eta \Sigma^{-}$                               
      & $ K^{0} \Xi^{-}$                      \\[\MySep] 
    \hline
    $ \pi^{-} \Lambda$                \rule{0em}{2.75\MySep}  
      & $\frac{1}{3} \DxD$
      & $- \TanXXX \DxF$
      & $\TanXXX \DxF$
      & $- \ROSixthHalf (\DxD + 4 \DxF + 3 \FxF)$
      & $- \frac{1}{3} \DxD$
      & $- \ROSixthHalf (\DxD - 4 \DxF + 3 \FxF)$ \\[\MySep]
    $ \pi^{-} \Sigma^{0}$
      &  
      & $\FxF$
      & $\frac{1}{3} \DxD$
      & $\CosXLVhalf (\DxD - \FxF)$
      & $\TanXXX \DxF$
      & $- \CosXLVhalf (\DxD - \FxF)$             \\[\MySep]
    $ \pi^{0} \Sigma^{-}$
      &
      & 
      & $\FxF$
      & $- \CosXLVhalf (\DxD - \FxF)$
      & $- \TanXXX \DxF$
      & $\CosXLVhalf (\DxD - \FxF)$               \\[\MySep]
    $ K^{-} n$
      &
      &
      &
      & 0
      & $- \ROSixthHalf (\DxD - 4 \DxF + 3 \FxF)$
      & $- \frac{1}{6} (\DxD + 3\FxF)$            \\[\MySep]
    $ \eta \Sigma^{-}$ 
      &
      &
      &
      &
      & $\frac{1}{3} \DxD$
      & $- \ROSixthHalf (\DxD + 4 \DxF + 3 \FxF)$ \\[\MySep]
    $ K^{0} \Xi^{-}$ 
      &
      &
      &
      &
      & \MinSepRule
      & 0
  \end{tabular}
\end{table} 

\newpage

\end{document}